\documentclass[preprint2]{aastex}
\usepackage{hyperref}
\usepackage{natbib}

\newcommand{\hip}{\emph{Hipparcos\,}}  
\newcommand{\kohlname}{Arnold Kohlsch\"utter } 

\shortauthors{Sandage, Beaton \& Majewski} 
\shorttitle{Mount Wilson Spectroscopic Parallaxes of Subgiants}

\begin{document}
\title{COMPARISON OF HIPPARCOS TRIGONOMETRIC AND MOUNT WILSON SPECTROSCOPIC PARALLAXES FOR 90 SUBGIANTS THAT DEFINED THE CLASS IN 1935}

\author{Allan Sandage,\altaffilmark{1,2}
Rachael L. Beaton,\altaffilmark{1,3}
\and Steven R. Majewski\altaffilmark{3}
}
\altaffiltext{1}{The Observatories of the Carnegie Institution of Washington, 813 Santa Barbara Street,
Pasadena, CA 91101}
\altaffiltext{2}{deceased}
\altaffiltext{3}{Department of Astronomy, University of Virginia,
Charlottesville, VA 22904-4325 (rlb9n, srm4n@virginia.edu)}

\begin{abstract}A history is given of the discovery between 1914 and 1935 of stars of intermediate 
 luminosity between giants and dwarfs with spectral types between G0 to K3. The Mount Wilson
 spectroscopists identified about 90 such stars in their 1935 summary paper
 of spectroscopic absolute magnitudes for 4179 stars. 
Called ``subgiants''
 by Str\"omberg, these 90 stars defined the group at the time. 
The position of the Mount Wilson subgiants in the HR
 diagram caused difficulties in comparisons of high weight trigonometric parallaxes being measured in the 1930s and with
 Russell's prevailing evolution proposal, and critics questioned
 the reality of the Mount Wilson subgiants. 
To show that the 1935 Mount Wilson subgiants are real, we compare, star-by-star, 
 the Mount Wilson spectroscopic absolute magnitudes of the 90 stars 
 defining their sample against those absolute magnitudes derived from \hip trigonometric
 parallaxes. 
We address concerns over biases in the Mount Wilson calibration sample and biases 
 created by the adopted methodology for calibration. 
Historically, these concerns were sufficient to discredit the discovery of subgiants in the Mount Wilson sample.
However, as shown here, the majority of the
 Mount Wilson stars identified as subgiants that also have reliable \hip
 trigonometric parallaxes do lie among the subgiant sequence in the \hip HR diagram.
Moreover, no significant offset is seen between the $M_V$ brightnesses derived from the Mount Wilson
 spectroscopic parallaxes and the $M_V$ values derived from \hip trigonometric parallaxes
 with $\sigma_\pi/\pi < 0.10$, which confirms in an impressive manner the efficacy of the original Mount 
 Wilson assessments.
The existence of subgiants proved that Russell's contraction proposal for stellar evolution from giants to the main sequence
 was incorrect. Instead, Gamow's 1944 unpublished conjecture that
 subgiants are post main-sequence stars just having left the main
 sequence was very nearly correct but was a decade before its time.
\end{abstract}


\section{Introduction}\label{sec:intro}
Often, in events leading to the beginning of a new field within
science, there appears an observation or experiment, not
understood at the time, that in the clarity of hindsight opened the field.
There are many such episodes. A famous one in physics is the
observation of at least 400 years ago that two bodies of
different weights, when released from a height, reach the ground
at the same time. This eventually lead to the general theory of
relativity.

In astronomy, the initial discovery of subgiant stars is such an episode. 
Subgiant stars are intermediate in luminosity in the HR diagram between the main-sequence 
 dwarfs and the giants near $M_V = +0.5$. 
The subgiants occur at absolute magnitudes between
 $M_V$ of $+2$ to $+4$ and spectral types between G0 and K3. 
Had they been fully understood from hints of their existence in the 1920s and
 their definitive discovery by \citet{Stromberg30} 
 in his absolute
 magnitude calibration, the path to an understanding of stellar
 evolution might have been hastened.

The purpose of this paper is to recount the discovery of subgiants and
 to point out the difficulty that their existence posed for
 Russell's early theory of stellar evolution. 
We then address the concern of early doubters of their existence, but 
 show that Str\"omberg's 1930 subgiants were real by comparing the Mount Wilson spectroscopic 
 parallaxes of \citet{Adams35} with modern \hip  
 trigonometric parallaxes for the $\sim$90 stars that defined the stellar class in 1935.

 The plan of the paper is this. 
 A history of the discovery of subgiants from 1917 to 1955 is given in the next subsection (Section \ref{sec:thedisc}). 
 This is an extension of the very brief account given in 
 \citet[][hereafter SLV03]{Sandage03}
 The dilemma of the existence of subgiants for Russell's proposal 
  \citep{Russell14,Russell25a,Russell25b} 
  about the direction of stellar evolution and Gamow's conjecture for a
  solution is discussed in Section \ref{sec:prevat}.

 Section \ref{sec:vatconf} is an account of the 1957 Vatican conference where the
  HR diagram of 25 subgiants with trigonometric parallaxes greater
  than 0\farcs052 was discussed. 
 The dilemma posed at that conference that field subgiants exist fainter than those in M67 is solved in
  Section \ref{sec:fieldsubgiants} with an account of the discovery 
  of the true ages of the old open clusters NGC\,188 and NGC\,6791.
  
 In Section \ref{sec:comp}, the spectroscopic parallax technique is introduced and
  \hip trigonometric absolute magnitudes 
  are compared with the 1935 Mount Wilson spectroscopic magnitudes of the $\sim$90 proposed 
  Mount Wilson subgiants.
 In Section \ref{sec:a35sub}, the biases in the \citet{Adams35} dataset are isolated
  and properly removed from the data, to address and quantify the historical criticisms 
  that challenged and colored perceptions of the veracity of the Mount Wilson spectroscopic dataset. 
 Section \ref{sec:c2sum} is a summary of the work presented here.

\begin{figure}
\includegraphics[width=\columnwidth]{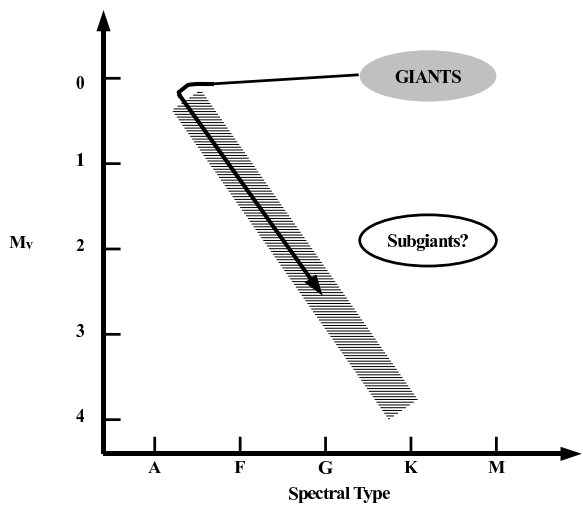}
\caption{\label{fig:russellevol} A schematic representation of Russell's evolutionary proposal for populating 
the main sequence by contraction of the giants and subsequent evolution
down the main sequence at nearly constant radius after the main
sequence stars become ``rigid." 
The subgiants could not be accommodated by this contraction scenario.
The position of the subgiants is adapted from the position of the subgiants in A35 (their Figure 1).}
\end{figure}

\begin{figure*}
\includegraphics[width=\textwidth,angle=90]{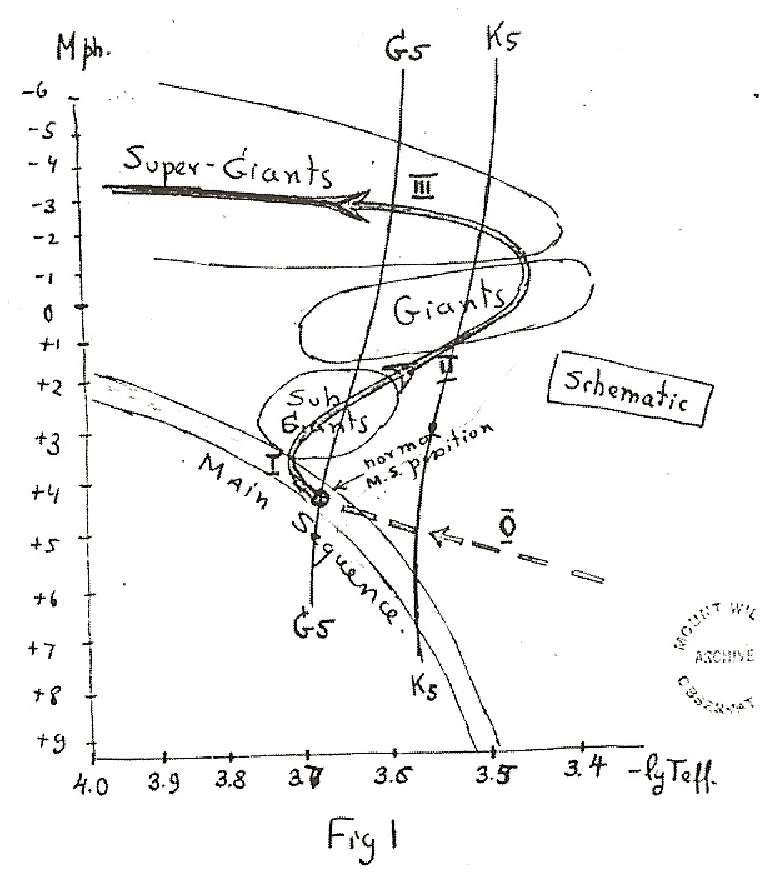}
\caption{\label{fig:gamowconj} Gamow's unpublished conjecture from a letter to Adams on
March 8, 1944 of evolution off the main sequence, through the
subgiants, and then to the giants.
This figure is reproduced from the Adams papers archive
at the Huntington Library, San Marino, CA. A more accessible
secondary source is \citet{Devorkin06}.}
\end{figure*}

\subsection{The Discovery}\label{sec:thedisc}

Following the invention of the method of measuring absolute
 magnitudes from stellar spectra by \citet{Adams14},
\citet{Adams17} published their first long list of absolute
 magnitudes for 500 stars of spectral classes from F0 to M. 
The distribution of absolute magnitudes for various intervals
 of spectral type clearly showed the separation of giants and
 dwarfs, discovered by \citet{Hertzsprung05, Hertzsprung07} and \citet{Russell14},
 but also showed a very few stars near $M_V = +2$ in the spectral
 interval from G0 to K3. 
The result is well shown in the distribution of \citet{Adams14}. 
A matrix representation of their 500 star
 sample was set out by \citet{Adams20} three years later,
 reproduced for easier access in \citet[][hereafter
 S04, Figure 15.6]{Sandage04}. 
 This representation clearly illustrated how very few of these intermediate
 luminosity stars were included in the magnitude-limited sample of
 \citet{Adams17} --- so few that one could question the reliability of their existence.
However, \citet{Curtis22} had summarized all spectroscopic absolute
 magnitudes known at the time, and again, with imagination,
 one could make out a continuum distribution of absolute
 magnitudes between $M_V = +0.5$ and $+5$ for spectral types G5 and K0
 (S04, Figure 15.5), yet the case for the existence of subgiants remained far from definitive.

\begin{figure}
 \includegraphics[width=\columnwidth]{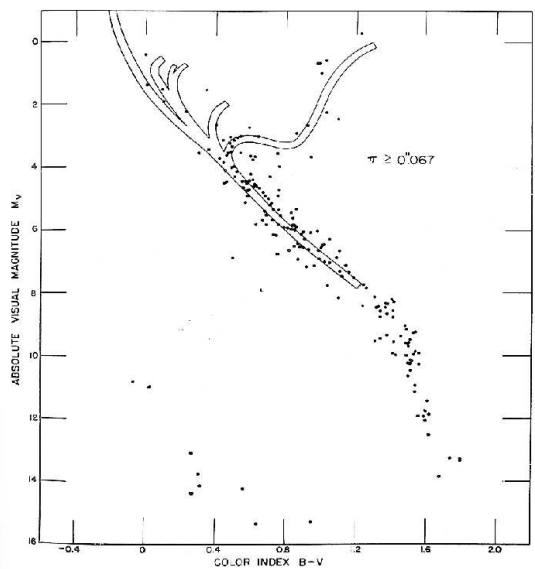}
\caption{\label{fig:vathr} The HR diagram for stars within 20 parsecs of the Sun
with relatively high weight trigonometric parallaxes known in
1957. Nine subgiants are fainter than the M67 subgiants, contrary
to the expectation of the Vatican Conference, which raised 
questions of their reliability. Diagram from 
\citet[][his Fig.\ 3, p.\ 295]{Sandage58b}.}
\end{figure}

\begin{figure}
\includegraphics[width=\columnwidth]{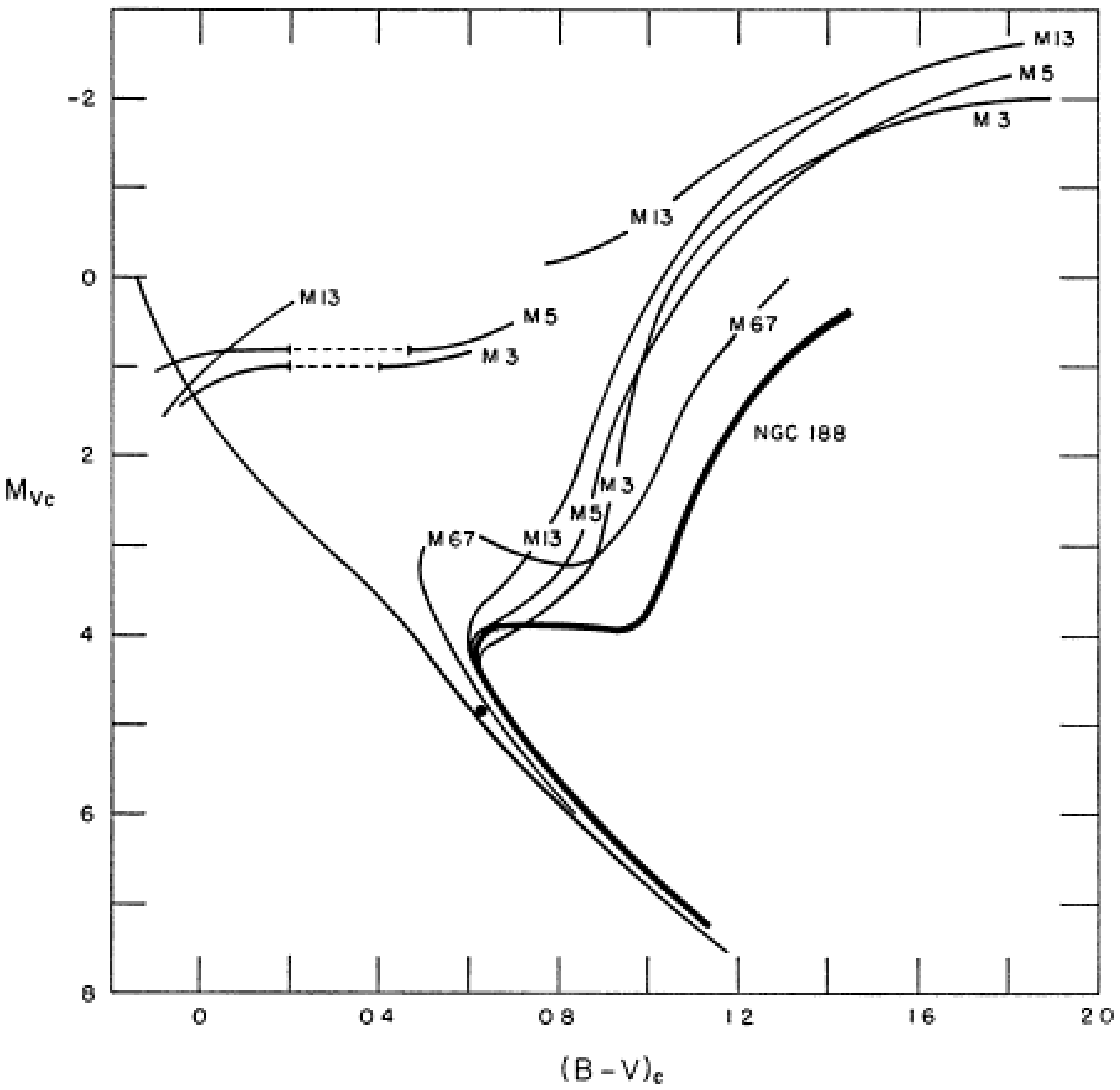}
\caption{\label{fig:n188} Solution to the problem of the Vatican Conference with
the discovery of the older cluster NGC 188, whose subgiant
sequence encompassed all known trigonometric field subgiants
known at the time. Diagram from \citet[][his Figure 10)]{Sandage62}.}
\end{figure}

 Subgiants were also consistently appearing in studies of a
  different kind where absolute magnitudes were estimated by
  independent methods. 
 The method of ``proper motion statistical
  parallaxes" had been invented by Jacobus C.~Kapteyn and \kohlname  
  in application to K stars, but remained unpublished until its application
  to spectroscopic parallaxes was realized by \citet{Adams14}.  
 It relied on the size of the proper motion for ``normal" stars, assuming that, statistically, in a
  large group of such stars, those with smaller proper motions were at larger distances. 
 \citet{Luyten22} used the method with 4446 stars of known proper motion and showed in a matrix, much like
  that used by \citet{Adams20}, that a continuum exists for stars of this type in the
  HR diagram between giants and dwarfs for spectral types G0 to K3. 
 But again, no such continuum exists later than K3. 
 Still, the number of such stars was minuscule and was generally ignored. 
 The work with the so called ``reduced proper motions" was
  repeated by \citet{Lundmark32}, again using more than 4000 stars.

The final list of the Mount Wilson spectroscopic parallaxes
 and the resulting HR diagram derived was published in the famous 
 summary paper by Adams, Joy, Humason and Brayton in 1935. 
In \citet[][A35 hereafter]{Adams35}, the subgiant sequence was unmistakable between
 G2 and K3 and at $\langle M_V \rangle = +2.5$, based on the statistical parallax
 calibrations of \citet[][ see S04, Figs.\ 20.2, 20.3]{Stromberg30,Stromberg32,Stromberg36}.
Adams et al.~wrote, ``The existence of a group of stars of
 types G and K somewhat fainter than normal giants has been
 indicated by the statistical studies of \citet{Stromberg30,Stromberg32}.
Although these stars may not be entirely separated from
 the giants in absolute magnitude, there is some spectroscopic
 evidence to support the suggestion."

During the later part of this Mount Wilson activity (1920-1935), such stars began appearing in
 lists of high weight trigonometric parallaxes and their existence could not be denied.
By 1936, six extreme subgiants with large trigonometric
 parallaxes could be said to define a subgiant sequence
 independently of spectroscopic parallaxes. 
These were $\mu$ Her (type
 G5, trigonometric parallax = 0\farcs119, $M_V = +3.75$); $\delta$ Eri (K0, 0\farcs111,
 +3.68); 31 Aql (G8, 0\farcs059, + 3.97); $\beta$ Aql (G8, 0\farcs073, +2.83);
 $\gamma$ Cep (K1, 0\farcs072, +2.223); and $\eta$ Cep (0\farcs070, K0, +2.68).\footnote{
 Values for the trigonometric parallax and $M_V$ are those from that era, but these values 
 are remarkably close to those found by the \hip satellite.}
The first three are fainter than the others by more than a
 magnitude, which shows the rather large intrinsic dispersion for the class. 
In the modern development of the \hip HR diagram, this
 intrinsic spread is seen to be about 3 magnitudes \citep{Perryman95,Kovalevsky98}.

In a historical and singular paper, 
 \citet{Morgan37} acknowledges the existence of subgiants, calculating the
 surface gravity of $\beta$ Aql, $\eta$ Cep, and $\gamma$ Cep
  --- the stars defining the subgiant class at that time --- 
 to be intermediate between the surface gravities of giants and dwarfs. 
It is also in this paper that W.~W.~Morgan sets out the case for abandoning the assignment of
 absolute magnitudes to the Mount Wilson spectroscopic two
 dimensional classifications, and replacing them by a continuum of numbers ranging from $-9$ to $+21$. 
The numerical continuum, 
  while ultimately related to the absolute magnitudes of the Mount Wilson system, 
  was uncalibrated and represented the raw line ratios measured from the spectra, 
  although intent was expressed to calibrate these values to absolute magnitudes at a later time.
 
This continuum of numbers was, eventually, replaced in the spectroscopic atlas of \citet[][hereafter MKK]{Morgan43} with 
 discrete boxes called luminosity classes, defining the MKK two dimensional system that
 discards the original Mount Wilson continuum system and replaces it by a
 digitized system with large classification boxes. 
In the 1943 MKK Atlas, the subgiant class was defined by the three stars used 
 by \citet{Morgan37} to which $\mu$ Her and $\delta$ Eri were added. 
These five stars are among the ``about 90 [stars] of types G and K somewhat fainter than normal giants''
 referred to by A35. 
The Yerkes subgiant class was designated as luminosity class IV and called subgiants following 
 the name first given by Gustaf Str\"omberg.

By 1955, Eggen could discuss a group of 20 bright subgiants,
 some of which have companions permitting calculation of their
 masses. 
The result was that the mean mass is near 1.2 M$_{\odot}$,
 consistent with the expectation that the stars in Eggen's 1955 list
 are ``evolving either from or toward the main sequence, near $M_V = +3.0$." 
\citet{Eggen55} showed that most of his 20 stars are within the
 borders of the subgiant sequence in the old open cluster M67
 defined by \citet{Johnson54}, although five are fainter ($\lambda$ Aur,
 $\beta$ Hyi, $\mu$ Her, 31 Aql, and $\delta$ Eri). 
Johnson had set up a photoelectric sequence in M67 in preparation for a complete
 photographic durchmusterung \citep{Johnson55}, but had
 published his sequence before the photographic work was complete.
In 1955, it was not clear how the subgiants tied onto the
 main sequence, a vital aspect that was sought but not clarified from
 previous work. 
Eggen's list of 20 subgiants are among those of A35 given in Table 2.1 here.

By 1957, Eggen produced a color-magnitude diagram
 \citep[Fig. 1 in][]{Eggen57} that clearly tied the few subgiants in a sample of
 275 field stars brighter than $M_V = 5.0$ to the M67 main sequence and
 to the base of the first ascent giants. 
Eggen's Figure 1 was
 mentioned by Walter Baade at the 1957 Vatican Conference \citep{Baade58}.
At that conference one of us could produce a list of 25 subgiants
 with high weight trigonometric parallaxes that defined the subgiant
 sequence discussed there \citep[][reproduced here as Figure \ref{fig:vathr}]{Sandage58b}.

     After the conference, two massive papers on motions, masses,
and luminosities of bright local field subgiants by \citet{Eggen60,Eggen64} 
completed the discovery phase for the subgiant luminosity class.

\section{ The Dilemma for Russell's Evolution Proposal; Gamow's Supposition}\label{sec:prevat}

\subsection{Russell's Dilemma}\label{sec:russell}
As stated in the Introduction, in his discovery paper and
 thereafter, 
  \citet{Russell14,Russell25a,Russell25b} 
 proposed that stars evolve by contraction from an initial giant phase,
 whereupon reaching the main-sequence dwarfs they become
 ``rigid,'' stopping the contraction. 
Although in the 1920s, Sir Arthur Eddington had proved that main-sequence stars do not 
 become ``rigid" (main-sequence stars have a perfect gas equation of state throughout),
 nevertheless Henry Norris Russell's several modified proposals remained
 highly influential throughout the late 1920s and early 1930s. 
The Russell proposal was widely discussed. 
An example is the Russell-evolution shown in the HR diagram by 
 \citet[][his Figure 30]{tenBruggencate27}
 or in the more accessible place in S04 (Chapter 17; Figure 17.1)

If the position of the subgiants in the HR diagram of A35 
 was correct as a sequence independent from that of the giants and dwarfs,  
 the Russell proposal could not be correct. 
Figure \ref{fig:russellevol} demonstrates the dilemma caused by the 
 positioning of the Adams et al. subgiants relative to the giants. 
The mean magnitudes and spectral types of the giants and subgiants in Figure \ref{fig:russellevol} are 
 taken from the A35 summary diagram (their Figure \ref{fig:russellevol}). 
Clearly, either Russell's evolution proposal was
 wrong, or subgiants did not exist.

Concerning attempts to show that subgiants did not exist,
 consider a most curious paragraph in Adrian Blaauw's scientific
 autobiography written in 2004, long after the existence of
 subgiants had been established \citep{Blaauw04}.  
As a beginning astronomer at Leiden in the late 1930s, 
 Adrian Blaauw had studied Gustaf Str\"omberg's papers
 and had concluded that ``[the observational data on proper motions]
 could equally be represented without the subgiant branch."\footnote{See also very similar 
 comments to this effect --- but instead explicitly critiquing R.E. Wilson's work rather than Str\"omberg's --- 
 in the 1979 interview of Blaauw by David DeVorkin, 
 available in the collection of oral histories on line by the American Institute of Physics
 ({\url http://www.aip.org/history/ohilist/5002.html}).}
This is an astounding conclusion in view of the increasing evidence
 from high weight trigonometric parallaxes and the accumulating
 evidence from spectroscopic parallaxes even at the time of 
 Blaauw's graduate work (the 1930's). 
  Later, in an extensive review of the calibration of luminosity criteria, 
  Blaauw was unable to reconcile the A35 sequence with the still
  small sample of subgiants having reliable 
  geometric distances \citep[see \S 2.14 of ][in particular his Fig.~5,
  which will be discussed in more detail below]{Blaauw63}.
Perhaps, Adrian Blaauw had not realized that the Mount Wilson
 subgiants of 1935 were linked with those of M67 in
 1955, or with the \hip trigonometric parallaxes of 1994,  
 given that the A35 subgiants were two magnitudes brighter than, and not connected to, the main sequence.
It is the purpose of this paper to show that the two groups of stars
 are identical by identifying the 90 Mount Wilson stars in a distinct 
 sequence near $M_V  = +2$ with \hip subgiants.

\subsection{Gamow's Conjecture}\label{sec:gamow}

   In a prescient letter from George Gamow to Walter Adams written in March
1944 and discovered by \citet{Devorkin06} in the Adams papers of the
Huntington Library archives, Gamow proposed a radically different
scenario of evolution off the main sequence caused by hydrogen
shell burning, shown in Figure \ref{fig:gamowconj} as track II. The subsequent
expansion of the radius accommodates the subgiants and giants, and
then a contraction (track III) gives the Wolf-Rayet stars, and
thence to the white dwarfs. Gamow's diagram is remarkably close to
the current evolutionary scenario. His conjecture was driven
by the results of the model builders, of which he was the most
tenacious. \citet{Devorkin06} tells the story primarily from the
many attempts at developing the theory in the early 1940s, all of which failed
in important ways.

The breakthrough supporting Gamow's conjecture did come in
 the early 1950s, largely resulting from improved observational precision
 that clearly showed a subgiant sequence attached to the main sequence.
Moreover, early evolution away from the main sequence was demonstrated
 in the Population I old Galactic clusters, M67 being the first in 1955.
In these new data, the subgiants fit naturally into the observational picture. 
Gamow's conjecture had been essentially correct, but the breakthrough was 
 most simply displayed before the eyes of the observers with their accurate, 
 faint color magnitude diagrams that began to be produced en masse after 1952.

\section{The 1957 Vatican Conference and Beyond}\label{sec:vatconf}

\subsection{The Vatican Conference}\label{sec:theconf}

   Walter Baade's population concepts were discussed from many
directions in an important conference in one of the first Semaine
d'Etude sponsored and held at the Vatican in 1957 \citep{O'Connell58}. 
Among the many topics discussed were comparisons of the HR
diagrams of Populations I and II relative to the color-magnitude
diagram of M67.

  To understand the discussion following a report by one of
us on the color-magnitude diagram of stars within 20 parsecs of the Sun
based on high weight trigonometric parallaxes, reproduced here in
Figure \ref{fig:vathr} \citep[][Figure 3, p.\ 295]{Sandage58b},
we must understand the mindset 
at the conference concerning the age dating of the globular
clusters, in particular the identification of the oldest Galactic clusters. 
A supposition, incorrect as it turned out, had been set by a report by one of us
 during the first days of the conference where a composite HR
 diagram of clusters of different ages, including M67, was
 compared with that of the old globular cluster M3 \citep[][Figure 3, on p.\ 42]{Sandage58a},
Here the main-sequence turnoff point for M3 is
 placed at nearly the same magnitude as that for M67.

     Because globular clusters were acknowledged to be the oldest
 objects in the Galaxy, and because, by the aforementioned
 agreement of the M67 turnoff point with the M3 termination
 magnitude by incorrect alignments in the HR diagram, it was believed
 at the conference that the M67 subgiants would define the locus
 of the faintest field subgiants. 
But with the clear observational
 violation of that supposition by the faintest subgiants in Figure \ref{fig:vathr}, 
 especially seen by 31 Aql, $\mu$ Her, and $\delta$ Eri, something
 did not fit, hence the extended and lively discussion recorded in 
 the conference report \citep[][pp.~299-302]{O'Connell58}.

     The culprit was, of course, the detail of the placing of the turnoffs of M\,67 and M\,3 
at the same magnitude in the aforesaid diagram. 
But it must be remembered that in 1957 the dependence of the main sequence
 on metallicity had not yet been discovered \citep{Sandage59},
 nor had the older galactic clusters NGC\,188 and NGC\,6791 been analyzed for their ages. 
The positioning of the main sequence according to metallicity has its own
 long history and is not especially germane to the subgiant story,
 but the discovery of clusters older than M\,67 is relevant.

\begin{figure}
\includegraphics[width=\columnwidth]{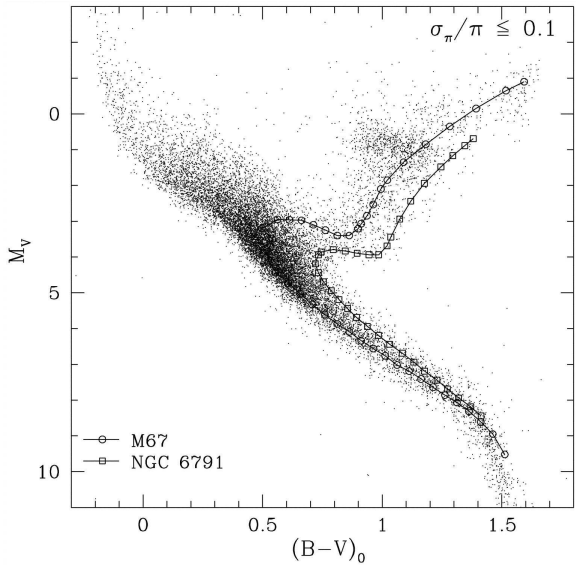}
\caption{\label{fig:slv03} The HR diagram for \hip field subgiants compared
with those for the old galactic clusters M\,67 (circles) and NGC\,6791 (boxes). Many
subgiants exist fainter than those in M\,67 and NGC\,188. NGC\,6791
provides a good lower envelope for the population. Diagram from SLV03 (their
Figure 12).}
\end{figure}

\subsection{Many Field Subgiants Exist Fainter than those in M\,67, NGC\,188, and NGC\,6791}\label{sec:fieldsubgiants}

Following the suggestions of Ivan King and Sidney van den Bergh, 
 photometry of NGC\,188 showed the cluster to
 have a turnoff luminosity fainter than M\,67 \citep{Sandage62}.
The placement of the M\,67 and NGC\,188 color magnitude diagrams as it was proposed in
 1962 is given in Figure \ref{fig:n188}. 
This shows that the NGC\,188 main-sequence turnoff is considerably fainter than that of M\,67.\footnote{
In Figure \ref{fig:n188} note that the turnoffs of the three
globular clusters are again placed at the NGC\,188 turnoff
magnitude. This is as incorrect as the placement of M\,67 at the
Vatican Conference. Hence, even as late as 1961 the position of
the main sequence as a function of metallicity 
 was not taken into account. A more correct diagram is in \citet[][Figs.\ 5  and 6]{Sandage86}.}

\citet{Kinman65} showed that the galactic cluster NGC\,6791
 was even older than NGC\,188. Many color magnitude diagrams have
 since been produced, perhaps the most accurate being that of \citet{Kaluzny95}.
 \citet{Chaboyer99} give a comprehensive review, 
 which is brought up to date by SLV03.
     The importance of NGC\,6791 is that it does define the lower
envelope of the field subgiant distribution. There are many subgiant stars
fainter than those M\,67, 
 but none statistically fainter than the subgiant sequence in NGC\,6791,
 as demonstrated in Figure \ref{fig:slv03} (adapted from Figure 12 of SLV03).

\section{Comparing the Spectroscopic and Trigonometric Parallaxes}\label{sec:comp}

The purpose of the paper is to show that the $\sim$90 subgiants
 discovered by the Mount Wilson spectroscopists in the late 1920s
 and early 1930s, as calibrated from statistical parallaxes by
 \citet{Stromberg30}, 
 are truly subgiants according to \hip 
 trigonometric parallaxes.
As an initial demonstration, we first compare the full A35 spectroscopic sample
 to the corresponding \hip trigonometric database to highlight
the general correlation between the two methodologies. 
Then, we will explore the key differences between the 
 A35 and \hip parallaxes and potential reasons for these
 differences. 
 However, before proceeding with these analyses, we recount briefly the historical 
 and physical premises that give foundation to
the spectrocopic parallax method as employed by the Mt. Wilson astronomers.

\subsection{The Mount Wilson Method of Spectroscopic Parallax}\label{sec:specpidef}

A comprehensive history of the Mount Wilson spectroscopic parallax program, work related to the project, 
 and biographical sketches of the contributing scientists, is related in S04; the following 
 is a synopsis of that account, 
 but with some further explication of details relevant to the present story.
The Mount Wilson sidereal spectroscopic program was one of the two ``key'' programs
 initiated on the 60$''$ Mount Wilson telescope upon its commissioning in 1908.
The sidereal program hoped to target all stars with proper motions
 and was comprised of two parts: 
 first, the measurement of radial velocities from 
 prism spectroscopy with a slit to improve the wavelength resolution, 
 and second, spectroscopic classification from these same spectra.
With these data in hand, George Ellory Hale hoped that 
 ``real progress would be made for stellar evolution'' (S04). 
The initial intent was to apply the then well-established Draper classification system,
 but the spectra, which spanned the range from $\lambda$4200 \AA~ to $\lambda$4900 \AA~,
 were capable of reaching a resolution of 16 \AA~mm$^{-1}$ \citep{Adams14} and
 proved difficult to impossible to characterize following the established Harvard system,
 which had been devised for spectra of a similar wavelength range but at a three times lower resolution. 
Even direct inspection of the plates by the Harvard team themselves failed to produce classifications
 consistent with those in the Draper catalog, 
 which demonstrated that the Draper system of spectral classification 
 was somewhat dependent on spectral resolution.
 
Thus, Walter Adams set out to ``measure'' spectral types from line ratios
 in lieu of ``estimating'' them from comparison spectra.
First attempts were fruitful, but continued to produce offsets of order
 $\sim$2 classification bins, producing tension between the Mount Wilson and Harvard efforts.
In attempting to resolve the tension, \kohlname first noticed that discrepancies in the
 line ratios were correlated with proper motion, 
 which was a clear indication that a luminosity effect drove differences in the line ratios. 
This observation was the ``seed'' for the \citet{Adams14} work that demonstrated the necessity
 of special consideration for the well established dwarf and giant luminosity classes 
 and, in turn, set the stage for the development of the technique of spectroscopic parallax.
Soon after the 1914 publication, the attempt to ``measure'' stellar spectral 
classes directly from line ratios was abandoned (and a form of the Draper system adopted). 
 Instead, realizing a spectroscopic distance technique was capable of building a large catalog of 
   absolute magnitudes without limitations of geometric parallax,
  the Mount Wilson astronomers changed their focus to the spectroscopic parallax technique.
This change of focus ultimately resulted in the A35 database of 4179 stars twenty years thereafter.

The initial luminosity criteria and an adaptation of the Harvard classification
 system to the higher resolution Mount Wilson spectra were published in a set of four papers
 presented to the National Academy of Sciences in 1916
 that provided the full details of the techniques employed as of that date
 \citep{Adams16a, Adams16b, Adams16c, Adams16d}. 
 The luminosity criteria were later expanded to both earlier and later stellar types, 
 and such stars were incrementally added to the total A35 sample \citep[most notably,][]{Adams20, Adams21, Adams26}.
Since the publication of the A35 catalog the luminosity criteria for spectral
 classification have evolved, but many of these early line ratios remain fundamental
 to luminosity classification \citep[see discussions by][and references therein]{Gray09}.
 
 It was noted in \citet{Adams21} that spectral types F through G exhibited
 a ``continuous change of absolute magnitude with line intensity'', 
 which contrasted with the ``marked discontinuity'' in the resulting magnitude distribution for the K and M types,
  which separate into distinct giant and dwarf classes within the solar neighborhood.
\citet{Stromberg30,Stromberg32} later isolated the subgiant luminosity class as the entity that
 connected the well known dwarf and giant sequences
 for the F and G spectral types using absolute magnitudes derived from statistical parallax.
Not all of the spectral lines associated with these classes exhibited a strong continuum of line strengths;
 the most important for the subgiants 
 were $\lambda$4077\AA, $\lambda$4215\AA, $\lambda$4324\AA, and $\lambda$4454\AA.
The lines $\lambda$4077\AA~ and $\lambda$4215\AA~ from Sr II 
 were identified as being sensitive to surface gravity as early as 
\citet{Adams16b}.\footnote{Adams note that these lines were 
 ``brighter in the spark than in the arc'' with respect to spectra
  obtained of the solar limb and upper chromosphere compared to the solar surface.} 
 The line $\lambda$4454\AA~ from Ca I is only used for types F8 to G in the detailed
  classification notes given by \citet{Adams21}.
 The Cr I line $\lambda$4324\AA~ is not discussed specifically until A35 (see their Table 1).
 The well-established empirical sensitivities to surface gravity 
  for the majority of the luminosity sensitive lines were given a physical underpinning by \citet{Saha21},
  who vividly illustrated the effect of electron number density 
  in populating the ionized metal states.
 As atomic theory progressed, a series of papers led by Walter Adams connected
  the parallel revelations produced via empirical study of stellar atmospheres 
  and those by pure theory \citep{Adams28, Adams29, Adams33}, and,
  in particular, highlighted the ability of observational astronomers to infer the physical
  conditions of stars long before there was a theoretical framework. 
 Most notable is the work of \citet{Adams28}, which combined the physical measurements
  of stars from binary observations by \citet{Russell14} with the 
  Mount Wilson line-ratio technique to conclude that it was feasible to 
  determine the physical conditions of stellar atmospheres from 
  appropriately chosen line ratios, more specifically 
   those line ratios conforming best to the theoretical predictions of line strengths
   from the Saha equation and were, effectively, unaffected by physics yet to be fully discovered. 
  
 While the electron-density dependence of the Saha ionization equation is, 
  indeed, the underpinning theory for why ionized metal lines are enhanced in 
  stars of lower surface gravity, this was not the only physics affecting the four lines
  used for exploration of the subgiants.
 In fact, two of the four lines used by A35 have additional properties that make them
  even more sensitive to subtle changes in surface gravity --- 
  the Sr II lines at $\lambda$4077\AA~  and $\lambda$4215\AA~ are resonant lines
  that arise from the ground state of the atom.
 Many of the other line ratios utilized by A35 (see their Table 1) are
  meta-stable levels of Fe II and Ti II, whose lower states are forbidden by
  quantum selection rules. 
 The reason for the particular sensitivity of $\lambda$4324\AA~ of Cr I and $\lambda$4454\AA~ of Ca I
  to surface gravity and, in consequence, their utility for the early identification of subgiants 
  is due to enhanced collisional broadening with in the atmospheres of dwarf type stars. 
 Modern spectral classification takes these, and other, detailed effects into 
  consideration.\footnote{For more information, we refer the interested reader 
  to the textbook on the subject by \citet{Gray09}.}
  
 Modern spectral classification can also exploit extremely high resolution spectroscopy,
  which, when compared to synthetically generated spectra, permits
  an even more nuanced portrait of stellar atmospheres.
 In particular, this body of work has made more apparent the underlying flaw of the 
  spectroscopic parallax approach:
  it remains unclear whether, given the numerous variables that affect stellar luminosity
  all stars of a given spectral class and luminosity type {\it do} have the same absolute magnitude 
 \citep[in particular, the effect of microturbulence is still an area of active exploration, e.g.,][]{Jaschek98}.
 Stated differently, the true intrinsic dispersion 
  of a given stellar class and luminosity type remains poorly constrained.
 To evaluate such a quantity, one requires sufficiently many stars with high weight trigonometric
  parallaxes to fully populate a sample for each spectral class and luminosity type ---
  something that is not fully feasible for even the \hip sample
  \citep[see discussion in][]{Jaschek98}, given the need for 
  obtaining complementary moderate resolution spectroscopy 
  and hiqh quality photometry for a large number of stars in each classification bin.
 A full understanding may be first be feasible with the combined astrometric, photometric
  and spectroscopic datasets provided by {\it GAIA} in the near future 
  (more than a century after the start of the Mount Wilson spectroscopic program).  
 In the end, the notion of ``measured'' luminosity classes following the Mount Wilson criteria
  was abandoned 
  by the astronomical community for the reasons already presciently argued by 
  \citet{Morgan37} that ``unreconcile-able difficulties of nomenclature'' would occur with such a measured system,  
  whose reduction curves would be highly sensitive to the specifics of the spectral data
  and analysis techniques employed for a given study.
  While the concept of ``measured spectral typing'' was ultimately abandoned,
  it is the subject of this work to evaluate whether the distances inferred via the spectrscopic parallax technique where 
  in themselves useful for advancing the understanding of stellar evolution with respect
   to the aforementioned problems with ``aging'' stellar sequences.
 
\begin{figure*}
 \includegraphics[width=\textwidth]{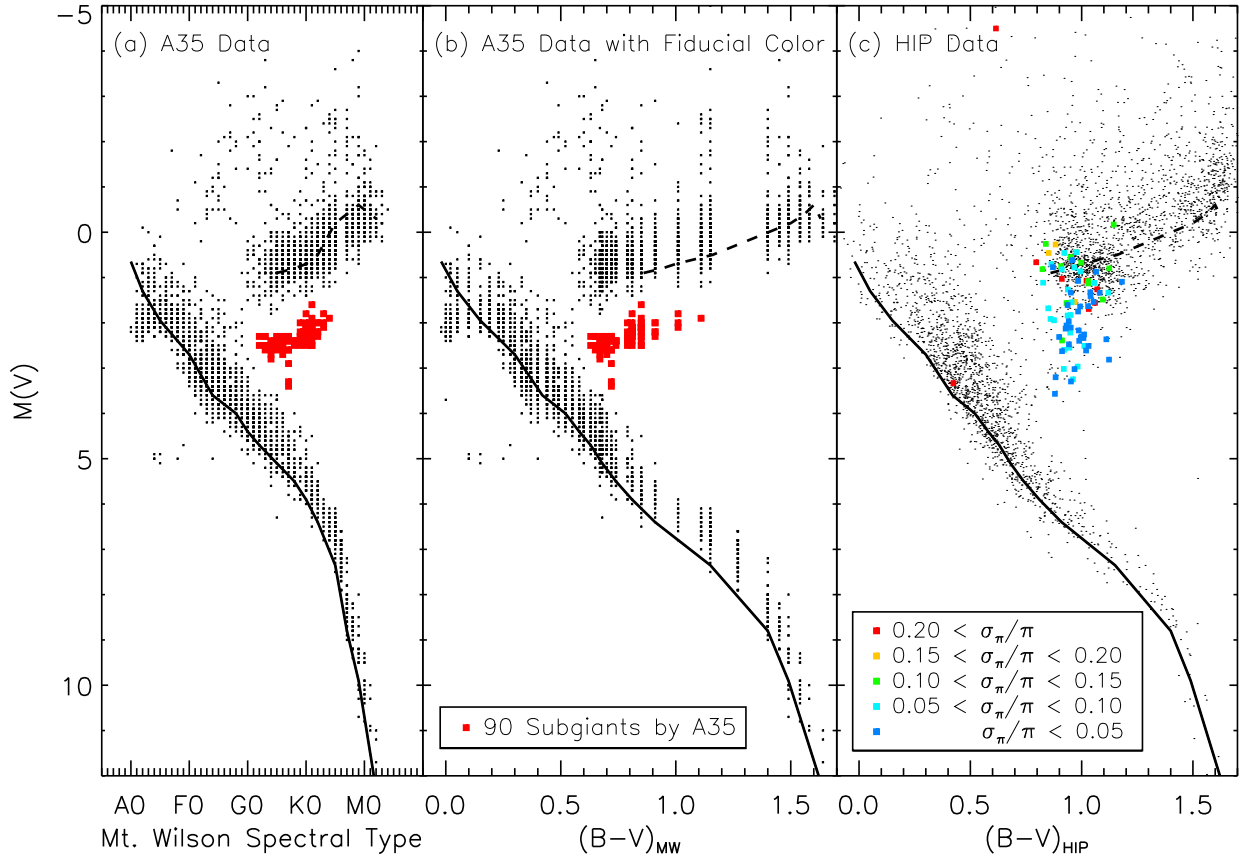}
\caption{ \label{fig:hr} The HR diagram for the 90 Mount Wilson subgiants from A35
in panel (a) plotted by spectral type
 and in panel (b) by conversion to a fiducial color, compared to (c) the
 \citet{vanLeeuwen07a,vanLeeuwen07b,vanLeeuwen08}
 \hip data for the same stars from
 the listings in our Table 1.  The solid line is an example main sequence and the
 dashed line is an example red giant sequence, with absolute magnitudes taken
 from \citet{Johnson66} and colors from \citet{Schmidt-Kaler82}. 
 Red filled squares in panels (a) and (b) indicate those stars we take as those
  identified as subgiants by A35.
 In panel (c) these same stars are color coded by their fractional
  parallax error ($\sigma_{\pi}/\pi$).
 This figure demonstrates that the bulk of the 90 stars identified in A35 are in fact subgiants.}
 \end{figure*} 
 
\subsection{The A35 Subgiants in the Context of the H-R Diagram}\label{sec:hr}

 To test whether the Mount Wilson-identified subgiants are real, we employ the 
  trigonometric parallax and photometric magnitude data contained in the 
  revised \hip catalog
  \citep{Perryman95,vanLeeuwen07a,vanLeeuwen07b,vanLeeuwen08}.
 Matches between the entire Mount Wilson list of 4179 stars as compiled by 
  A35 to the \hip catalog were made solely through
  the star names --- predominantly the Henry Draper number, available for
  90\% of the stars --- as identified in the A35 catalog.\footnote{
     Because an electronic version of the extensive Mount~Wilson data table, 
      spanning some 93 pages of the Astrophysical Journal, does not exist, 
      these pages were scanned into digital form and then passed through 
      optical character recognition software (OCR) to compile output tables.  
     The OCR software required considerable training because the fontface and
      typesetting styles used in the 1935 print journal are not included in standard OCR templates. 
     The output digital tables were then reformatted and checked manually for errors.
    Detailed data for only the subgiants are presented directly in this paper, but 
      a digital version of the full A35 database will be made available online.}
 We present the full matched catalog in the HR diagrams of Figure \ref{fig:hr}. 
 We note that the A35 catalog contains spectral types, apparent magnitudes, absolute magnitudes
  and the derived spectroscopic parallax, whereas the \hip catalog
  contains apparent magnitudes and trigonometric parallaxes. 
 It is also important to bear in mind that the colors and apparent magnitudes in the
   \hip catalog come from a diverse set of ground-based sources, 
   but we find the data to be reliable and homogeneous enough to make our point
   (i.e., even with whatever systematic and random uncertainties may lie within the
   catalog, the correlations we find are strong).
 In Figure \ref{fig:hr}a we reproduce the HR diagram from the original A35 paper (their Figure 1) 
  based on our digitalization of their original catalogue while in
  Figure \ref{fig:hr}b we convert the A35 spectral type to a fiducial $B-V$ color  
  using the standard observed values defining each spectral class as presented in \citet{Schmidt-Kaler82}. 
 In Figure \ref{fig:hr}c, we convert the \hip trigonometric parallaxes
   to absolute magnitudes using the apparent magnitude
   information provided in that catalog.

 The stars considered to be subgiants are not explicitly identified in A35.
 Instead, the authors refer to ``stars of type G and K somewhat fainter than normal giants"
  and a mean magnitude of $M_V \sim 2$. 
 We used the A35 description to identify their subgiants as those stars having 
  $1.6 < M_V < 3.4$ and spectral types from G2 to K4 in the A35 tables;
  these stars are highlighted as red filled boxes in Figures \ref{fig:hr}a and \ref{fig:hr}b. 
 Fortunately, the majority of the stars we take to be the A35 subgiants 
   are clearly separated from the red giant and main sequences in Figure \ref{fig:hr}a, consistent
   with the A35 description, and there are few ambiguous examples. 
 We identify 90 such stars as likely to be the A35 ``subgiants", and this
  matches well their statement that they see ``about 90 some stars" meeting their criteria.

 In Figure \ref{fig:hr}c these same stars are highlighted and color-coded by their \hip fractional 
  parallax error, $\sigma(\pi)/\pi$,
  which is a standard measure of the reliability of trigonometric parallaxes. 
 The majority of the stars highlighted in Figures \ref{fig:hr}a and \ref{fig:hr}b 
  remain ``subgiant-like'' in the \hip HR diagram of Figure \ref{fig:hr}c, 
  but the clean separation between the subgiants, giants and main sequence
  in the A35 data is no longer present;
  instead they form the more familiar continuous evolutionary sequence
  from the main sequence to the red giant branch.
 We emphasize that no \citet{Lutz73}
 corrections have been applied to the \hip data,
  but that such corrections are on a scale that will not affect the overall impression given in 
  Figure \ref{fig:hr}c.
 We discuss the potential effects of the Lutz-Kelker bias in more detail in Section \ref{sec:LKE}.  

 Table 1 presents the relevant A35 and \hip data for the 90 stars identified as 
  subgiants in Figure \ref{fig:hr} (the red boxes). 
 Most of the columns of Table 1 are self explanatory. 
 Column 9 supplies the value of $\sigma(\pi)/\pi$ from \hip, a useful
  parameter guiding the interpretation of results.
 The crucial columns are 10 and 11 where the Mount Wilson absolute
  magnitudes determined from the spectra are compared with the
  \hip trigonometric values.
 Absolute magnitudes from the \hip data are derived by combining the 
  trigonometric parallaxes and apparent magnitudes, while the values adopted by the Mount Wilson
  astronomers are given directly in the catalogue of A35.\footnote{In particular, A35 states that
   their subgiant stars were ``selected based on the strength of 
   the lines 4077, 4215, 4324 and 4454. Special reduction curves
   based on trigonometric parallax were used to determine the 
   absolute magnitudes of these stars."}

 Despite the remarkable correspondence between the spectroscopic 
  and trigonometric absolute magnitudes in Table 1, 
  there are two pronounced differences.
 First, unlike what is seen in the modern \hip panel (Figure \ref{fig:hr}c), 
  the subgiant sequence as positioned by Adams et al. (Figures \ref{fig:hr}a and \ref{fig:hr}b) 
  was neither attached to the main sequence on the left or onto 
  what in later HR diagrams was the base of the first ascent giant branch on the right.
 Second, the overall scatter along each sequence in the A35 diagram (Figures \ref{fig:hr}a and \ref{fig:hr}b)
  is smaller than --- or at least on par with ---
  that of the more accurate, modern magnitudes in the corresponding 
  \hip diagram (Figure \ref{fig:hr}c).
 This second effect is most dramatic for the subgiant sequence in A35 (colored points in Figure \ref{fig:hr}),
  which is nearly horizontal and spans only $\sim$ 1 magnitude of $M_{V}$.

 Given these morphological differences in the subgiant branch,
  the significance of the subgiants for stellar evolution were not realized at that time.
 If it had been, however, the placement of these stars could not be explained by
  the prevailing Russell giant-to-dwarf stellar contraction scenario (Figure \ref{fig:russellevol}).
 In contrast, subgiants are a natural phenomenon for the evolutionary 
  model proposed with Gamow's conjecture (Figure \ref{fig:gamowconj}).
 More specifically, 
  the position of the subgiants in Figure \ref{fig:hr} is easily explained 
  as stars transitioning from the main sequence to the first-ascent giant branch after hydrogen core exhaustion.
 As summarized above, this understanding of the role of subgiants in stellar evolution,
  in particular as relative age indicators for single stellar populations,
  was not to be fully conceptualized until the early 1950s.
 
 Nevertheless, even if it was not fully appreciated or accepted
  at the time, the Mount Wilson spectroscopists had, in fact, discovered the subgiant 
  sequence from their spectral line classification method. 
 Meanwhile, from independent considerations, Gamow had
  proposed the correct evolution scenario, 
  but it was a decade before his unpublished conjecture was 
  substantiated based on observations of the main-sequence turnoffs  
  and subgiant branches in a number of galactic and globular clusters in the 1950s and 1960s.

\begin{figure*}
  \includegraphics[width=\textwidth]{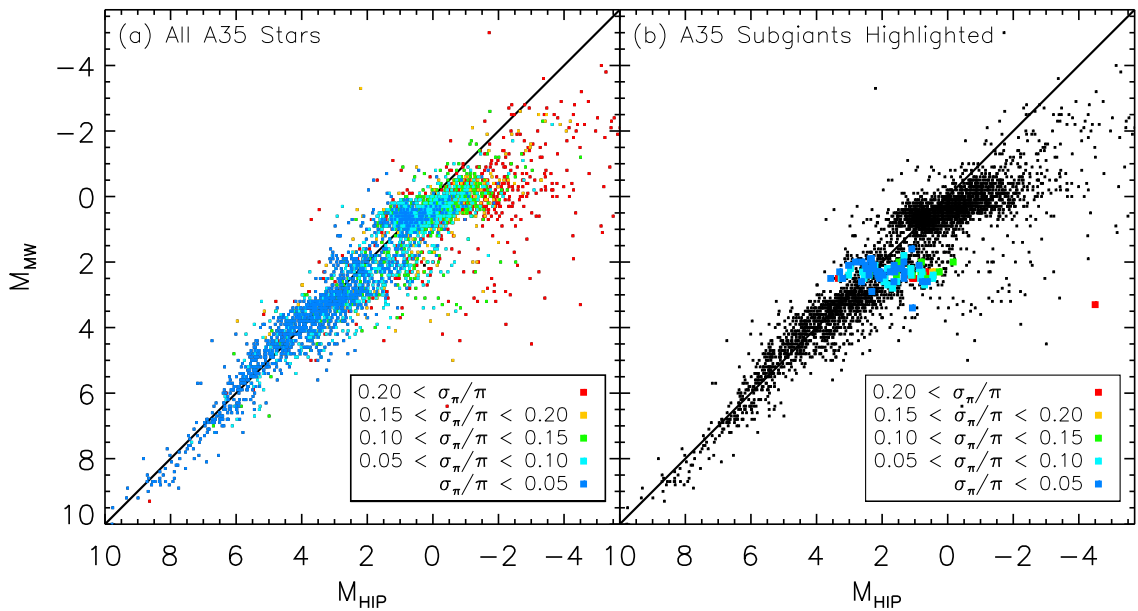}
  \caption{ \label{fig:abscomp}Comparison of Mount Wilson spectroscopic absolute
    magnitudes --- in panel (a) for the full A35
    sample and in panel (b) highlighting the 90 subgiants ---  with \hip
    absolute magnitudes for the same stars. 
    The color coding in both panels is by the fractional trigonometric parallax error,
     or $\sigma_{\pi}/\pi$.
    The A35 subgiant absolute magnitudes are incredibly compressed (spanning only $\sim$ 1 magnitude)  
     compared to those of \hip (spanning $\sim$ 4 magnitudes), both of which are given in Table 1.}
\end{figure*}

\subsection{Magnitude Comparisons}\label{sec:magcomp}

 Although the visual comparison of the Mount Wilson HR diagram and CMD 
  (Figures \ref{fig:hr}a and \ref{fig:hr}b, respectively)
  with the corresponding \hip CMD (Figure \ref{fig:hr}) is quite illustrative, 
  we directly compare the two sets of absolute magnitudes for all stars in the Mount Wilson tables 
  in Figure \ref{fig:abscomp}.  
 Remarkably, a strong correlation is found among all luminosity classes and spectral types.
 These results convey the striking efficacy of the Mount Wilson spectroscopists to get the true
  absolute magnitudes from spectra, and, in particular, for the previously unidentified subgiant class
  (those stars with $1.6 < M_V < 3.4$ and spectral type from G2 to K4).

 There are, however, some differences visible in Figure \ref{fig:abscomp}
  and these differences vary with luminosity class. 
 The main-sequence stars by far exhibit the cleanest correlation, 
   $\langle M_{V, {\rm Mt W}} - M_{V, {\rm \hip}}\rangle = 0.18 \pm 0.02$; this is
  largely because the A35 absolute magnitudes of these stars were calibrated
  using the trigonometric parallaxes available to the Mount Wilson astronomers at that time. 

 The subgiant and giant branches, on the other hand, 
 exhibit a correlation that is visibly shallower than 
  that for the main sequence. 
 Additionally, the mean offsets for the giants and subgiants are substantially larger: for the giants
   $\langle M_{V, {\rm Mt W}} - M_{V, {\rm \hip}}\rangle = 0.46 \pm 0.02$
 and for the subgiants
   $\langle M_{V, {\rm Mt W}} - M_{V, {\rm \hip}}\rangle = -0.24 \pm 0.07$. 
 From visual inspection of Figure \ref{fig:abscomp}, however, it is clear that those stars with
  the lowest $\sigma_{\pi}/\pi$ agree better with the A35 results (cyan and blue in Figure \ref{fig:abscomp}).
  If we restrict to $\sigma_{\pi}/\pi \le 0.05$, then we have, 
   for the main sequence, 
   $\langle M_{V, {\rm Mt W}} - M_{V, {\rm \hip}}\rangle = 0.06 \pm 0.02$,
   the giants, 
     $\langle M_{V, {\rm Mt W}} - M_{V, {\rm \hip}}\rangle = 0.15 \pm 0.04$,
   and the subgiants,
     $\langle M_{V, {\rm Mt W}} - M_{V, {\rm \hip}}\rangle = -0.12 \pm 0.02$.
  These comparisons are summarized in Table 2.2.
  Unlike those of the main-sequence stars, the absolute magnitudes of the evolved stars were 
   calibrated against a sample of {\it statistical} parallaxes, which may 
   explain the differences in the behavior of the evolved versus main sequence stars
   in Figure \ref{fig:abscomp}.
 
 Although the overall similarities between the A35 and \hip absolute magnitudes
  are strong, the magnitude differences are not insubstantial.
 Furthermore, that the severity of the magnitude 
  differences are correlated with luminosity class may help to explain
  why the A35 subgiant sequence was not appreciated 
  in the literature of the time.
 To explore these differences, we first consider the potential effects of the 
  Lutz-Kelker Effect in the \hip absolute magnitudes in Section \ref{sec:LKE} 
  as a potential source of the observed mean magnitude offsets in Figure \ref{fig:abscomp}.  
 In Section 5, we revisit the historical discussions of
  the A35 absolute magnitudes to reassess published criticisms
  of the A35 calibration technique.

\subsection{The Lutz-Kelker Effect}\label{sec:LKE}

 The Lutz-Kelker Effect (LKE, hereafter) is an insidious and complex bias 
  afflicting trigonometric parallax samples that was discussed in detail by \citet{Trumpler53} 
  and then parametrically formalized
  by \citet{Lutz73}.  
 The effect is defined as an offset between the mean absolute magnitudes for classes of stars
  as determined from trigonometric parallax samples and the true mean absolute magnitude
  for that stellar class.
 Because the effect is only identified in the mean values derived for a stellar class its effect on individual stars
  is difficult to intuit. 
 Here we first review the origin of the bias, and use a brief schematic
  analysis to assess the severity of the LKE in the matched A35-\hip sample.
 A more rigorous derivation of the relevant LKE parameters is given in Appendix A. 

 At the parallax limit of a sample, the volume element just outside of the limit is larger 
  than the one just inside.
 A symmetric error distribution applied to both volume elements will systematically scatter 
  more stars into a sample than out --- 
  because the outer (and larger) volume element is likely to have more stars
  (though the magnitude of the effect clearly depends on the underlying stellar distribution).  
 This asymmetrical scattering has the consequence that, statistically, more stars of a given
  stellar class will be scattered into a parallax-limited sample than out. 
 The parallax for a star scattered into the sample is obviously observed to be larger than the true value, 
  and by consequence, both the inferred distance and intrinsic flux are underestimated.
 Thus, the mean absolute magnitude of a particular stellar class within the sample will be offset compared 
  to the true intrinsic mean magnitude for that class; 
  this offset is defined as the Lutz-Kelker correction.
   
  Early numerical explorations of the bias 
  parameterized it by $\sigma(\pi)/\pi$, the ratio of the
   error in the parallax to the parallax itself \citep{Lutz73}.
   The bias is found to be significantly pronounced when $\sigma(\pi)/\pi \gtrsim 0.2$.
  Because the parallax error is nominally set by the systematics of the trigonometric parallax program
   (e.g., temporal baseline, detector pixel scale, overall flux sensitivity),
   the LKE grows in its relevance 
   towards the volume limit of the survey. 
  Though the LKE is defined as a statistical offset for a {\it sample} of stars of a given class, 
  \citet{Lutz73} argue that 
  their defined correction could be used to correct the derived absolute magnitudes for {\it individual} stars, 
  and it often has.
 More recently, the logic behind this practice and Lutz \& Kelker's original
  suggestion that the correction could
  be applied to individual stars has been challenged \citep{vanLeeuwen07a,Smith03,Francis14}.
 Regardless of this ``Lutz-Kelker Paradox'', it is nonetheless naturally the case that stars with a larger parallax
  error will show a larger error in derived absolute magnitudes, and, as a class, would be expected to exhibit a larger
  systematic offset; it is our goal to assess whether this phenomenon has any significant effect in our comparison
  of A35 absolute magnitudes to those derived from \hip parallaxes.
 In Figure \ref{fig:abscomp}a, individual stars are color coded by $\sigma_{\pi}/\sigma$. 
 As anticipated, those stars with higher 
  fractional trigonometric parallax errors, show, on average, the largest deviations from 
  the A35 magnitudes --- though the degree of this scatter is highly dependent on the luminosity class.  

 Is the LKE primarily responsible for 
  the magnitude offsets between A35 and \hip (Section 4.2; Table 2)? 
 The LKE as applied to the \hip catalog,
  has been explored by numerous authors, including, but not limited to, 
  \citet{Oudmaijer98}, \citet{Brown98}, and \citet{vanLeeuwen07a}.
 Generally, it is advised to reject all stars with fractional errors 
  greater than 10\% or $\sigma(\pi)/\pi > 0.10$,
  but a full treatment of the LKE, and any other relevant biases, 
  requires detailed modeling of the specific sample in question 
  \citep[e.g., ][]{Sandage02,vanLeeuwen07a,Smith03,Francis14}.  
 While the very form and magnitude of the LKE has also been reassessed and debated (see previous
  references) we adopt the formalism of \citet{Sandage02} for our calculations, which are primarily
  illustrative; because subsequent treatments of LKE
  tend to find bias corrections of an overall smaller magnitude than those of \citet{Sandage02}, our 
  examples can then be seen as an upper limit to the expected LKE influence on the \hip data.

 To place the magnitude offsets between the Mount Wilson and \hip data in context,
  we consider the magnitude and distance distribution for the A35 stars, for which 
  the faintest is $V=10.5$, the majority are brighter than $V=8$,
  and the most distant is only $d =500$ pc. 
 In comparison, \hip, is complete for m$_V$ = 7.3 to 9.0, reaches 
  a limiting magnitude of m$_V$ = 12.4,
  and has a median parallax precision of 0.97 mas (i.e., that of a 1000 pc star).
 As given in Table 2.2, the magnitude differences between A35 and \hip data
  are generally small compared to the precision feasible for the A35 magnitude measurements,
  especially so when restricted to $\sigma(\pi)/\pi < 0.05$.
 Using the $\sigma(\pi)/\pi < 0.05$ criterion, a star 
  with the median parallax error would have a corresponding distance ($d=50$ pc)
  well within the \hip sampling volume, 
  and at a magnitude for which the \hip sample is complete.
 Using this fiducial, we conclude that any trigonometric parallax bias offsets in our sample
  are going to be quite small, and unlikely to influence strongly the impression
  of the magnitude comparisons given in Section 4.2. 

 This rough assessment of the role of the LKE in our comparison suggests 
  that the LKE does not significantly influence the general impression given 
  in Figures \ref{fig:hr} and \ref{fig:abscomp}.
 In Appendix A, a detailed numerical exploration of the LKE 
  along the lines of \citet{Sandage02}
  supports the conclusion that LKE is not significantly affecting the \hip data for the A35 sample.

\section{Reconciling Criticisms of the A35 Subgiant Sequence}\label{sec:a35sub}

In the previous section we showed that the subgiants discovered in 
 the A35 dataset are bona fide subgiants in the \hip catalog. 
In our historical account of the discovery of subgiants, we have discussed in detail
  why the prevailing stellar evolution theory in 1935, the Russell contraction scenario,
  may have biased the immediate interpretation of the A35 results at the time of publication.
We have not, however, discussed the reasons why the A35 
  data, which show a clear subgiant sequence, were insufficient 
  to challenge Russell's hypothesis (Figure 1).
 
Numerous works investigated the reliability of the A35 database,
 as it remained one of the largest catalogs of absolute magnitudes well into 
 the middle of the 20th-century.
A full discussion of this body of work, as well as rigorous comparisons 
 to pre-\hip databases, is given by \citet{Blaauw63}.
Our goal, to show conclusively that the Mount Wilson spectroscopists discovered 
 subgiants in their seminal 1935 work, requires addressing the concerns expressed 
 for the A35 calibration methodology by revisiting the doubts on its reliability.
More specifically, we must determine  whether specific, cited 
 problems artificially created the subgiant sequence in Figure \ref{fig:hr}. 

 First, we briefly reiterate the methods employed by A35:
  \citet{Adams14} first detected a correlation between specific sets of line intensities 
  and the proper motion of the star --- 
  more specifically, that for a given spectral type the line intensities for
  small and large proper motion stars were systematically different.
 From this initial discovery, a series of line ratios were established as a means of identifying both 
  the spectral type and the absolute magnitude (luminosity) of a star from spectroscopy.
 This work became the basis for the two decade observing campaign that culminated 
  in the 4179 member catalog in A35.

 This method of ``spectroscopic parallax'' itself was never a point of contention in the literature; rather,
  it was the calibration of the technique against existing parallaxes that raised concerns.
 The calibration required two key decisions:
  (i) a choice of a suitable external calibration sample, and 
  (ii) the grouping of the existing spectral line data into appropriate classes.
 For the former, Gustaf Str\"omberg's parallax program
  was selected by A35 as an ideal calibration sample, even
  though his statistical parallax sample for giant stars suffered effects from Malmquist bias.
 Clearly, this bias would propagate into the A35 catalog, 
  but in manner difficult to intuit; concerns over this problem 
  led to a general distrust of the reliability of the absolute magnitudes for giant stars, 
  and a dismissal of the A35 subgiant sequence as likely attributable to the effect.
 For decision (ii), the Mount Wilson spectroscopists assumed that the observational
  scatter was driving the distributions (and not intrinsic differences between stars) 
  and thus chose to group the line ratios 
  based on the peaks in their distribution 
  --- a process that can suppress outliers (see Appendix B) --- 
  instead of using an independent criterion to classify the stars by spectral type and luminosity class
   that would yield a more representative scatter.
 The effect of this choice on the A35 catalog is less immediately obvious, 
  but, as will be explained in Section \ref{sec:a35corr}, it
  resulted in an overall suppression of the true observational errors in the A35 sample.
 This suppression is visible in Figures \ref{fig:hr}a and \ref{fig:hr}b as the narrow 
 vertical dispersion across all sequences (main, subgiant, and giant), 
 but most conspicuous for evolved star sequences in Figure \ref{fig:hr},
  and the overall ``disconnected'' nature of all three luminosity sequences in the HR diagram.

 Both of the above choices in the calibration process 
 by the Mount Wilson spectroscopists were heavily discussed in the literature, 
  and became a source of contention regarding the legitimacy of spectroscopic 
  parallaxes for any stars not on the main sequence, 
  i.e., any stars not calibrated directly to trigonometric parallaxes.
 Adrian Blaauw in his 1963 review of luminosity criteria comments 
  that the complex statistical arguments over the A35 calibration in the literature
  ``may have created the impression that these systematic errors are of an intricate nature
    and the Mount Wilson results therefore less useful 
    [than other absolute magnitude samples]'' \citep{Blaauw63}.
 With more geometric parallaxes in hand, Blaauw demonstrated the efficacy of the
  A35 data for the dwarfs and we repeat this analysis with \hip parallaxes
  to demonstrate the same for the giants and the subgiants.
 We first address the cited concerns over the effects of Malmquist bias in the selected
 calibration sample on the A35 
 survey in Section \ref{sec:malmquist}.
 Then we qualitatively describe the problems with the A35-adopted 
 calibration methods in Section \ref{sec:a35corr}
  and correct the A35 magnitudes for these effects in Section \ref{sec:correcting}. 
 A detailed quantitative derivation parallel to the more qualitative
 discussion in Section \ref{sec:a35corr} is given in Appendix B.

 \begin{figure*}
 \includegraphics[width=\textwidth]{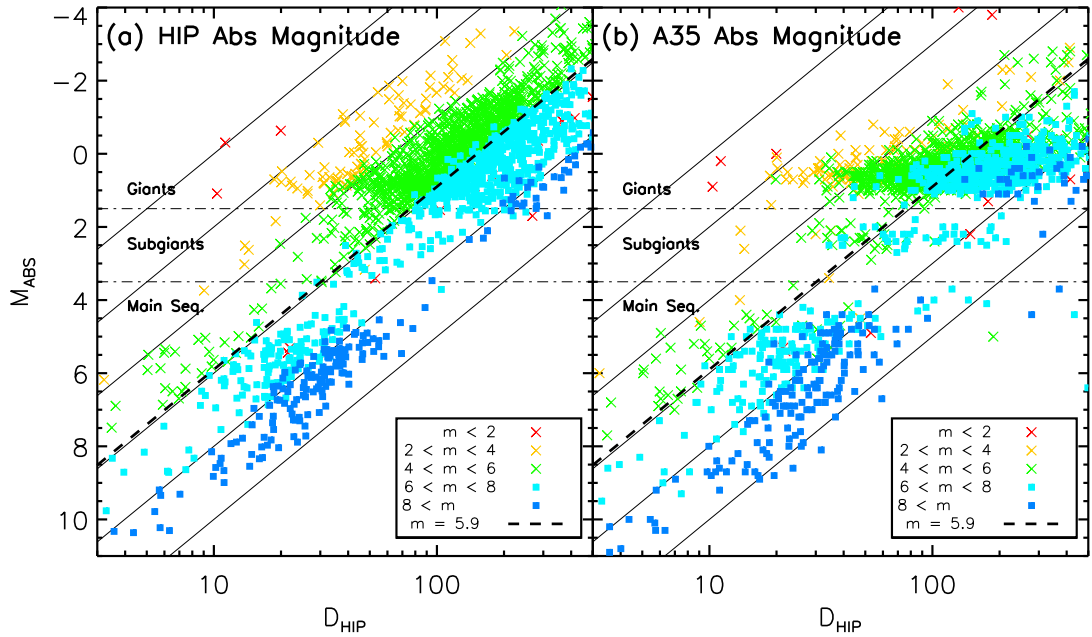}
 \caption{ \label{fig:spaenhauer} Spaenhauer diagrams for the matched (a) \hip 
    and (b) A35 magnitude samples, 
     both of which plotted using the \hip distances ($D_{HIP}$). 
   The thin black lines are the 
    apparent magnitude limits of 0, 2, 4, 6, 8 and 10 magnitudes.
   The horizontal dot-dash lines mark the separation between luminosity classes 
    of dwarf, subgiant, and giant, while the dashed line separates stars brighter (crosses) and
    fainter (squares) than $V=5.9$.
   The Malmquist bias has the effect of creating intrinsic luminosity trends with distance 
    appear at the faint limit of a sample. 
   As apparently fainter stars are added any trends
    imposed by the Malmquist bias will disappear, whereas real trends
    will persist with the fainter samples.
   Panel (b) shows that the subgiants persist as a distinct group beyond $V=5.9$.
    }
 \end{figure*}

\subsection{A Test for an Incompleteness (Malmquist) Bias in the 1930
              Calibration by Str\"omberg}\label{sec:malmquist}
 
 The A35 sample featured three luminosity classes cleanly distinguished by their position in the HR diagram: 
  dwarfs, subgiants and giants. 
 The absolute magnitudes were determined from spectral line ratios
  using independent calibrations for each luminosity class and spectral type. 
 More specifically, the main-sequence stars were calibrated with 
  absolute magnitudes from trigonometric parallaxes, 
  whereas the, on average, more distant subgiants and giants were calibrated to statistical parallaxes. 
 In Figure \ref{fig:abscomp}, the difference in calibration techniques is obvious, 
  with the Mount Wilson absolute magnitudes calibrated to the statistical parallaxes 
  showing a stronger deviation from the \hip values than those calibrated to trigonometric parallaxes.
 
 The statistical parallaxes used to calibrate the giant and subgiant sequences were 
  derived using a sample complete to an apparent magnitude of $V=5.9$ \citep{Stromberg30},
  whereas the full A35 sample contains significantly fainter stars ($V=10.5$).
 In a sample that is not volume limited, intrinsically luminous objects
  are systematically over represented because they can be surveyed 
  over a larger volume of space
  than intrinsically fainter objects at the same apparent magnitude limit.
 The \citet{Stromberg30, Stromberg32} calibration sample was subject to this ``Malmquist bias", 
  which skews the mean absolute magnitude of a sample systematically brighter than the true mean absolute magnitude. 
 Thus, with their samples calibrated with the Str\"omberg data, the A35
 absolute magnitudes calculated for the subgiants and giants sample would be 
  systematically brighter than those that would be calibrated against an unbiased sample. 

 The bias would also apply to any dwarf stars misidentified as giants and for which the giant
  star calibration were utilized.  
 Under one interpretation, the A35 subgiants could be stars misplaced 
  in the HR diagram and made systematically even brighter due to the Malmquist bias.
 Meanwhile, with the Malmquist bias in the giant star calibration skewing the magnitudes 
  of real giant stars systematically brighter,
  the separation between luminosity classes would be exaggerated as well, and presumably
  contribute to the disjoint appearance of the subgiants in the A35 HR diagram.    

 To judge the viability of these concerns, 
 we can diagnose the role of the Malmquist bias, and even treat its effects, 
  on the A35 magnitudes.
 The Spaenhauer diagram \citep[see review on the topic by][]{Sandage01}, an effective tool for detecting and correcting 
  the effects of the Malmquist bias, plots the absolute magnitude ($M$) 
  versus the distance to the object ($D$).
 In the Spaenhauer diagram, the presence of the Malmquist bias produces fictitious behavior in the magnitudes with 
  respect to distance. 
 A key requirement for diagnosing the Malmquist bias in a sample is 
  having distance estimates that are independent of the 
  luminosity estimates being explored. 
 One means to diagnose and assess the strength of the Malmquist bias
  is to assess how the implied behavior of data in the Spaenhauer 
  diagram changes as fainter samples are added incrementally \citep{Sandage01}.

 We demonstrate this technique in Figure \ref{fig:spaenhauer}, 
  where we plot (panel a) the \hip absolute magnitudes versus the 
  trigonometric distance, which at the maximum radius in this sample, $D < 400$ pc, 
  should not show any significant effects due to this bias.   
 The thick black dashed line represents the flux limit of the 
  Str\"omberg statistical parallax sample ($V=5.9$).
 For the stars with $m_V < 5.9$ (``x'' symbols) in Figure \ref{fig:spaenhauer}a, 
  those at the greatest distance (the giants) tend to have an intrinsically brighter absolute magnitude.
 Were one limited to only a $V<5.9$ sample, one would 
  {\it incorrectly} infer that there are no intrinsically faint stars beyond $40$ pc. 
 But when stars having fainter apparent magnitudes 
  are added (the filled boxes), the previously 
  apparent behavior --- i.e., that there are no intrinsically faint stars beyond $40$ pc --- 
  disappears at those distances at which the sample is now complete.
 While this particular demonstration is quite simple, in other datasets exploring more complex astronomical questions 
  the Malmquist bias can induce many false impressions \citep[see a historical review by][]{Sandage01}.

 Figure \ref{fig:spaenhauer}b is similar to Figure \ref{fig:spaenhauer}a except the Spaenhauer analysis
 is applied to A35 absolute magnitudes, which are also plotted 
 against the \hip distance ($M_{HIP}$).
 In Figure \ref{fig:spaenhauer}b we see the same behavior as in Figure \ref{fig:spaenhauer}a for the main-sequence 
  stars ($M > 3.5$); again, the magnitude trend with distance disappears as  fainter apparent
  magnitude samples are considered.
 If we restrict ourselves to those stars with $V > 5.9$, the giants and subgiants are nearly 
  horizontal in magnitude with respect to distance and the two branches are offset from one another.
 While both trends --- the horizontal shape and the separation between branches --- are shown to be specious
  by comparison to the pure \hip data in Figure \ref{fig:spaenhauer}a,
  we can also see by Figure \ref{fig:spaenhauer}b that the specious character is not introduced by the 
 Malmquist bias but that, in fact, these features are intrinsic to the A35 absolute magnitudes: 
 As we view increasingly fainter samples in Figure \ref{fig:spaenhauer}b, the shape of the giant and subgiant branches 
  and their separation do not change. 
 This demonstration effectively dispels the notion that the A35 sample for giants
  and subgiants was significantly compromised by the Malmquist bias
  in the calibration of \citet{Stromberg30}, 
  and that this bias does not readily account for a false introduction of the subgiant sequence
  nor does it fully account for the shape of the giant branch.
 For a full exploration of these criticisms and their implications in the literature of that time, 
  we refer the reader to \citet{Blaauw63}, 
  in which the Malmquist bias could
  not be fully tested owing to the still relatively small number of luminous stars with trigonometric parallaxes.

 \begin{figure*}
  \includegraphics[width=\textwidth]{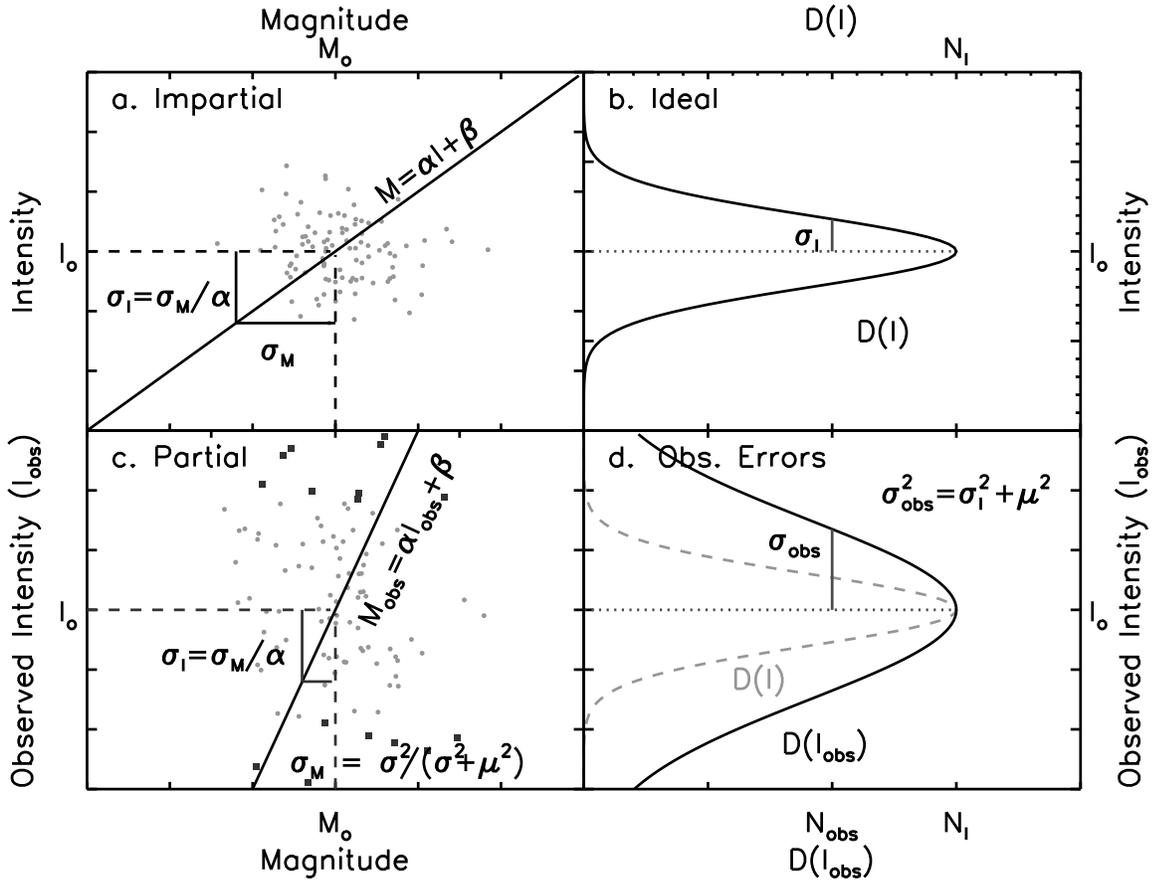}
  \caption{\label{fig:calibmeth} Demonstration of the impartial ({\it panels a and b}) and
   partial ({\it panels c and d}) calibration techniques for fictitious datasets without (i.e., ``ideal") 
   and with observational uncertainties, respectively. 
  In the left panels, a fake dataset is drawn from the Gaussian distributions (filled circles)
   shown in the right panels; in both cases the black solid line represents the intrinsic spread, $D(I)$, of 
   the data unconvolved with observational errors.
  For the case of the impartial calibration ({\it panels a and b}), all of the data drawn from the
   Gaussian distribution is used to derive the calibration line $M=\alpha I+\beta$ in {\it panel a}.
  For the case of the partial calibration, observational errors (presumed to be normally
  distributed with an RMS spread $\mu$) 
  inflate the observed spread 
  of line intensities, $D(I_{obs})$ ({\it dashed curve}), but some measurements ---
  the {\it black filled circles} in {\it panel c} ---
   are rejected from the calibration sample as ``extreme observational outliers". 
   The spread of the remaining sample ({it grey circles}) mimics that of the intrinsic, error-free sample, $D(I)$
  ({\it panel d}).
  However, the resulting calibration ({\it panel c}), $M_{obs}=\alpha I_{obs}+\beta$, 
   is ``torqued'' with respect to that one would obtain from the error-free dataset ({\it panel a}).  
  The actual quantitative differences are derived rigorously in Appendix B, 
   following the discussion by \citet{Trumpler53}.}
 \end{figure*}

\subsection{Correcting the A35 Calibration Method}\label{sec:a35corr} 

Having determined that the Malmquist bias is not the source of the morphological 
 differences between the A35 and \hip CMD, we now
 explore the A35 calibration as the culprit.
 For the A35 data, visual estimates of relative spectral line intensities were calibrated
   empirically to absolute magnitudes derived from trigonometric parallaxes for
   the dwarfs and to statistical parallaxes for the evolved stars \citep{Stromberg30}.
 In this section, we qualitatively describe the luminosity calibration used
  by A35 and explore how their process induced systematic errors in their derived absolute magnitudes.
 A complementary quantitative derivation of these errors 
 is given in Appendix B.

 Combinations of line intensities had already been shown to classify the stars by their spectral type
  and in \citet{Adams14}, the Mount Wilson spectroscopists discovered a set of line intensities that were 
  sensitive to luminosity. 
 Once the \citet{Stromberg30} sample of 1647 stars became available,
  the Mount Wilson spectroscopists began to explore how to exploit it for calibrating their data. 
 The \citep{Stromberg30} stars were first grouped into their spectral classes based on the Mount Wilson criteria.
 Within each of these classes, the line intensities were observed to cluster 
  around certain mean values ($I_o$), and these were correlated with the proper motion of 
  the star (a proxy for the distance to the star). 
 Each of these distinct line intensity groups represented one of the classes from the
  modern luminosity classification system, though at the time this
  was primarily limited to the dwarf and giant classes.
 However, from his calibration sample of 1930, Str\"omberg had already noted the existence of stars intermediate 
  in luminosity between the giants and dwarfs and he termed these stars ``subgiants'' 
  for their location relative to the giant branch. 
 Thus, for each of the {\it three} groups of line intensity, a calibration curve was determined as
  a linear regression of the form $M = \alpha I + \beta$
   \citep[example curves are given in][their Fig.~1]{Adams16a}.
 These calibrations, ultimately derived from the Str\"omberg sample of 1647 stars, were then applied to the
  full Mount Wilson sample of 4179 stars in the seminal 1935 catalog. 

 Though at a cursory glance this calibration procedure seems valid, 
  there is a key flaw in the method: 
  the line intensities were used {\it both} to determine calibration groupings {\it and} 
  to derive the calibration curves. 
 Because these two separate steps rely on the same underlying data, they are not independent (impartial).
 The calibration is, therefore, ``partial'' to those line intensities tightly clustered around the 
  visually identified mean value and this bias will have an effect magnitude distribution derived 
  from the calibration.
  Moreover, there is a clear assumption in this procedure that the intrinsic variation 
  within a given classification bin is small or at least well characterized by the 
  calibration dataset (which was not the case).
  Stated differently, the Mount Wilson spectroscopists attributed the visible scatter
   entirely to uncertainties of an observational nature, which effectively
   ignores any ``cosmic'' or intrinsic variation (which could be larger or smaller than the observational scatter).
 In the case of A35, the Mount Wilson spectroscopists proceeded as if both steps, grouping 
  and calibration within the group, were uncorrelated.
 This assumption produced both of the morphological differences between Figure \ref{fig:hr}b and Figure \ref{fig:hr}c:
  (i) the exaggerated separation between the luminosity classes, 
  and (ii) the suppressed dispersion along each of the three sequences.

 To illustrate the nature of the bias generated by this procedure,
  Figure \ref{fig:calibmeth} uses a pair of fictitious data sets to compare
  ``impartial'' and ``partial'' methods for which the grouping and calibration 
  steps are uncorrelated and correlated, respectively.
 Figures \ref{fig:calibmeth}a and \ref{fig:calibmeth}b 
 demonstrate the ideal or ``impartial'' calibration scenario, applied to an ``error-free" dataset.
 In Figure \ref{fig:calibmeth}a, we see the ideal calibration curve between absolute magnitude and line intensity
 derived from the distribution given in Figure \ref{fig:calibmeth}b. 
 In this case, intensity values ($I$) are selected by an ``impartial'' or independent means and the 
  distribution of values, $D(I)$, and its intrinsic, cosmical spread,
  $\sigma$, are well sampled.
 Here the calibration utilizes the linear fit $M = \alpha I + \beta$ 
  and the scale of the uncertainty in magnitudes, $\sigma_{M}$, follows the usual propagation, 
  $\sigma_{M} = \alpha \sigma$.
 Our description here is analogous to the \citet{Blaauw63} discussion of ``impartial'' calibration 
  of a dataset having no observational errors, but where 
  the spread in line intensities is only given by a true
  intrinsic variation among a stellar type.
  
 In Figure \ref{fig:calibmeth}c the linear calibration fit resulting from a ``partial'' calibration
  is ``torqued'' or rotated from that derived in an ``impartial'' manner.
 The origin of this torque is the ``partial'' nature of the adopted line intensity groupings.
 The observed spread of line intensities in a sample where random
  observational errors are known to be convolving
  with the intrinsic line intensity spread, $D(I_{obs})$, there is a natural tendency to reject 
  extreme outliers as being produced by the wings of the observational uncertainty distribution --- presumed itself
  to be normally distributed with a dispersion $\mu$ --- as a means to winnow the sample 
  to one more representative of the error-free distribution, $D(I)$.
 However, the resulting fitted calibration, $M_{obs} = \alpha I_{obs} + \beta$
  (based on those points shown in light gray, which fall in the distribution $D(I)$)
  not only has a steeper slope, 
  but also a much smaller dispersion in $M$ than would obtain in the error-free case (Figure \ref{fig:calibmeth}).  
 As shown in Appendix B, for the same line intensity spread, 
  $\sigma_i$, the $\sigma_M$ dispersion is reduced by a factor 
  of $\sim$ 1/$\sqrt(1 + \mu^{2}/\sigma^{2})$ from the partial to the impartial analysis.

 Thus, the effect of pursuing a ``partial" calibration is not only
  an improper calibration relation for each relevant luminosity class,
  but the resulting range of magnitudes is also artificially compressed 
  through the suppression of outliers.
 The former effect is evident in the comparison of 
  $M_{A35}$ to $M_{HIP}$ in Figures \ref{fig:abscomp}a and \ref{fig:abscomp}b, 
  in which the subgiant and giant branches appear ``torqued''
  with respect to a one-to-one correlation,
  as well as in the compression of the subgiant and giant sequences in Figures \ref{fig:hr}b and \ref{fig:hr}c.
 Thus, the origin of the key differences in the A35 and \hip CMDs lies
  in a subtle bias in the calibration technique used in A35.

 Now, we can ask a different question: 
 Though we have already shown that the subgiants in A35 are real, 
 was their isolation in A35 a fortunate accident of this calibration error?

\subsubsection{The Systematic Error}
The ``partial'' nature of the A35 magnitude calibration curves was discussed at 
  length in the literature of the time,
  most notably by \citet{Stromberg39}, \citet{Stromberg40}, \citet{Stromberg41},
  \citet{Russell38}, \citet{Russell40}, and \citet{vanRhijn39}.
 These authors not only described the nature of the calibration bias, but also 
  proceeded to correct the absolute magnitudes.
 The correction procedure relied on the existence of distances derived independent of A35,
  which are used to correct the ``torque'' shown in Figures \ref{fig:abscomp}a and \ref{fig:abscomp}b. 
 A more detailed history of these papers and their efforts to correct the A35 catalog 
  is given by \citet{Blaauw63}, including more modern use of the A35 catalog 
  by \citet{Oke57,Oke59} and \citet{Wilson57}.

 While the ``torque'' correction method worked well for the dwarfs of all spectral types, application
  to the more distant subgiants and giants remained difficult owing to the 
  smaller number of such stars having independently measured distances.
 As ground-based trigonometric samples grew, however, the correction method was applied 
  to more giant type stars. 
 For the A35 giants, however, the intensity line calibration method so torqued the axis 
  (see the ``squished'' giant branch in Figure \ref{fig:hr}a), that the resulting corrections to A35
  had large uncertainties.
 For the subgiants in Figure \ref{fig:hr}a, so compressed as to be completely detached from 
  the main sequence and giant branches, correction of the A35 magnitudes was deemed impossible.
 An example calibration to spectral type K is given in Figure 5 of \citet{Blaauw63},
  to which Blaauw states, 
  ``With the accidental errors so strongly dominating the cosmic dispersion, 
  it is obviously meaningless to try to introduce corrections to the Adams values for the giants.
  For the subgiants the situation is similar.''
 Thus, the A35 data for the giants and subgiant classes were largely ignored owing to the inability to 
  reconcile the bias introduced by the ``partial'' calibration. 

 It is important to note, that even at the time of Adrian Blaauw's 1963 summary article, 
  only 41 K-giants and 12 K-subgiants had well measured trigonometric parallaxes ---
  representing less than 10\% of the total number of such stars in A35 \citep[][S04]{Blaauw63}.
 With the substantially larger \hip database, 90\% of the A35 stars now have independently measured parallaxes.
 In the next section, we endeavor to use these data to correct the full A35 HR diagram for the calibration bias 
  to determine if the subgiants were created artificially by these biases.

 \begin{figure*}
  \includegraphics[width=\textwidth]{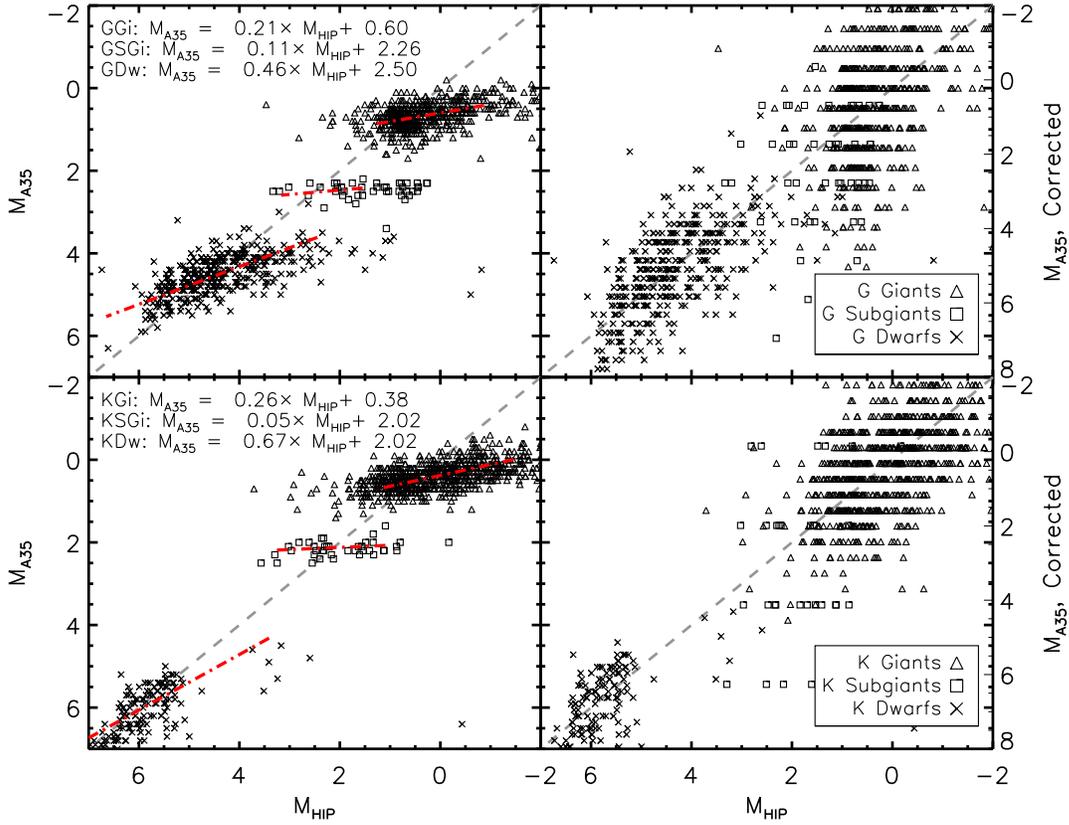}
  \caption{\label{fig:blaauwcorr}Procedure followed to calibrate the A35 magnitudes to the
    \hip magnitudes following the detailed discussion
    in \citet{Blaauw63}, and separated by G ({\it top}) and K ({\it bottom})
    spectral types.  The left panels display the original data
    with fits to the dwarfs (GDw, KDw), subgiants (GSGi, KSGi) and 
     giants (GGi, KGi).  We used the luminosity classifications
    in the \hip catalog. In the right panels these calibration fits
    are applied to the A35 magnitudes and refit to demonstrate the removal of the torque.}
 \end{figure*}

\subsubsection{Correcting for Systematic Errors in the Luminosity Calibration}\label{sec:correcting}

 In this subsection, we use the \hip trigonometric parallaxes 
 \citep{Perryman95,vanLeeuwen07a,vanLeeuwen07b,vanLeeuwen08}
  as a comparison sample to remove the torque in the A35 calibration curves (e.g., Figure \ref{fig:calibmeth}c to \ref{fig:calibmeth}a). 
 First, we will describe the calibration method utilized by previous authors.
 Then, we will demonstrate its effectiveness for the G and K type stellar classes 
  (i.e., those containing the majority of the subgiants in A35).
 Lastly, we apply the correction for all stars in A35 to revisit Figure \ref{fig:hr}
  and create a corrected version of Figure 1 from A35.
 With this we aim to answer the question:
  if we remove the calibration torque in the A35 sample, do the 90 stars in A35 remain subgiants
  or do they largely join the main sequence or giant branches?
 Effectively, we ask whether the ``partial'' calibration method --- the separation of stars
  into tightly concentrated fiducial values --- artificially create the 
  distinct, separated subgiant class? 

 To correct the A35 magnitudes using the \hip sample, 
  we transform our previous discussion into our set of variables specific to A35 and \hip.  
 Where possible, we follow the conventions used in the literature. 
 We separate the sample by the A35 spectral type: A, F, G, K, and M, and 
  then by luminosity class: dwarf, subgiant, and giant, using the A35 absolute magnitude. 
 We first compute the mean magnitude for the 
  spectral type and luminosity class in both the A35 sample, $\langle M_{A35} \rangle$, 
  and the \hip sample, $\langle M_{HIP} \rangle$, and define the difference between the two as,
 \begin{equation}
  \label{eq:magdiff}
  \Delta\langle M \rangle = \langle M_{HIP} \rangle - \langle M_{A35} \rangle. 
 \end{equation}
 Then, we fit a linear regression for each luminosity class. 
 The slope ($m$) of the fitted line represents the torque on the feature due to
  the improper calibrations (e.g, the relationship of Figure \ref{fig:calibmeth}c). 
 The quantity $A$, defined by \citet{Russell38} as the ratio of the dispersion of the measurement 
  and the true dispersion is empirically derived by: 
 \begin{equation}
  A = (1/m)-1 . 
 \end{equation}
 We then can compute the corrected magnitude for each star within a spectral and luminosity class using 
  the following equation \citep[][; Equation 14]{Blaauw63}:  
 \begin{equation}
  \label{eq:mcorr}
  M_{Corr} = M_{A35} + \Delta\langle M \rangle + A(M_{A35}+ \Delta\langle M \rangle - \langle M_{HIP} \rangle)
 \end{equation} 
 where $M_{Corr}$ is our final corrected A35 absolute magnitude, 
       $M_{A35}$ is the A35 absolute magnitude and,
       $A$ is a quantity calculated from the results of the linear regression for a
       spectral and luminosity class and defined in Equation 2. 

 The process for correcting the magnitudes is demonstrated in Figure \ref{fig:blaauwcorr} for the A35 subgiant
  spectral classes G (top panels) and K (bottom panels).
 In the left panels of Figure \ref{fig:blaauwcorr}, the original A35 magnitudes are plotted against 
  the corresponding \hip magnitudes ($M_{HIP}$, as in Figure \ref{fig:abscomp}) --- 
  and illustrate the severe torque on the subgiant and giant luminosity calibrations in these classes. 
 For each luminosity class, a linear regression is fit against those \hip 
  stars with $\sigma_{\pi}/\pi$ less than 0.10 for the giants and less than 0.05 for
  the subgiants and dwarfs. 
 The results of the linear regression are given in the left panels. 
 In the right panels, the results of the fits are used in Equation \ref{eq:mcorr} 
  and applied to $M_{A35}$. 
 In the right panels we accomplish both of the goals of this procedure: 
  not only has the torque been corrected, but the observational scatter
  is increased to appropriate levels. 

\begin{figure*}
 \includegraphics[width=\textwidth]{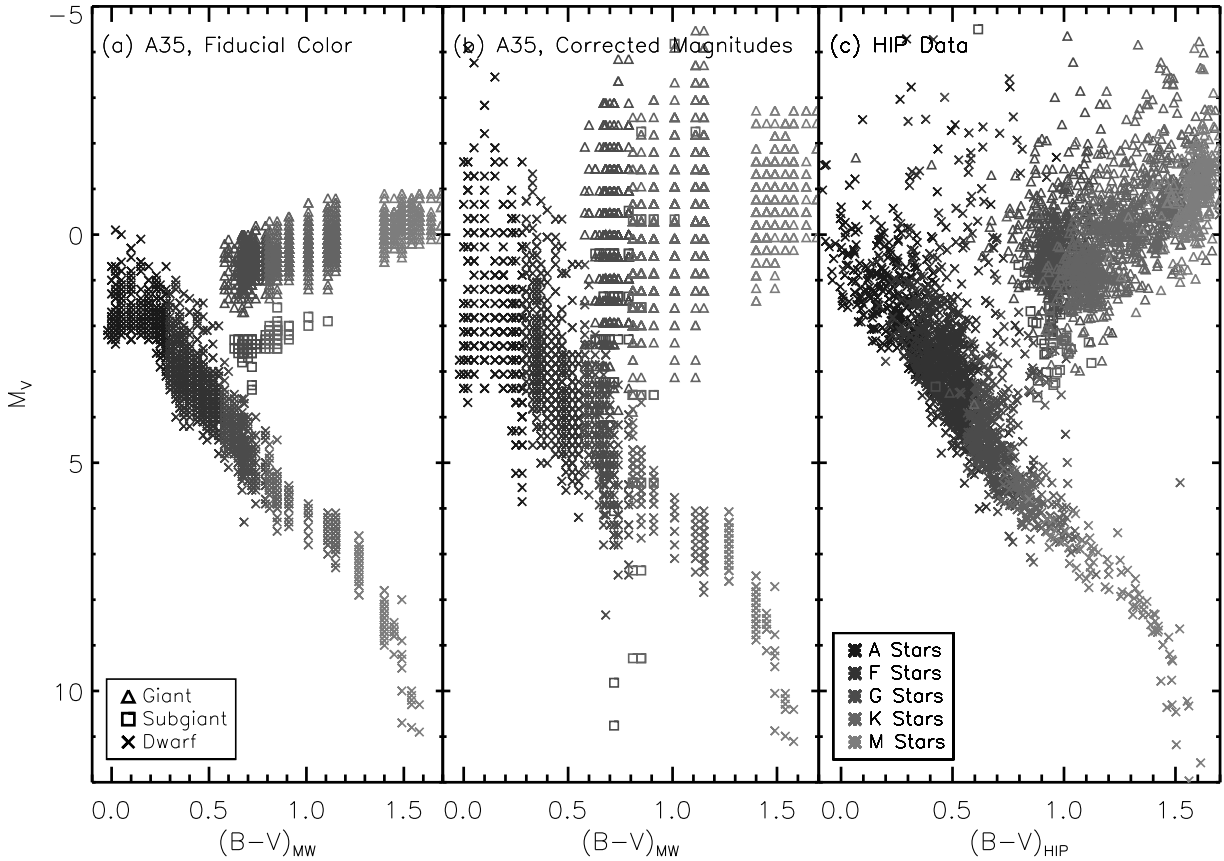}
 \caption{\label{fig:hrcorr} Application of the magnitude
    corrections to the A35 HR Diagram.
   In panel (a) we reproduce the A35 HR diagram with color axis translated
    from spectral types to a fiducial color.
   In panel (b) we apply the magnitude calibration procedure described in
    Section~\ref{sec:a35corr}, and in panel (c) we reproduce the
  \hip HR diagram for the A35 sample.
  Panel (b) demonstrates how correcting for the bias in the A35 methods
  brings visual impression of the A35 subgiants into better agreement
  with that of \hip, most notably the subgiants now connect
  the main sequence and giant branch. The magnitude scatter
  along the sequences is now greater than that of \hip.}
 \end{figure*}

 Having demonstrated that the larger \hip database can correct the A35 magnitudes 
  for the subgiants and giants, we proceed to use the \hip data to correct
  the entire A35 sample and create the true HR diagram for the A35 sample. 
 In Figure \ref{fig:hrcorr}a the original A35 HR diagram is reproduced with the 
  A35 spectral types color-coded in grayscale.
 In Figure \ref{fig:hrcorr}b the corrected A35 magnitudes are plotted
  and the scatter within each of the luminosity and spectral type bins is greatly increased. 
 In Figure \ref{fig:hrcorr}c the \hip HR diagram is shown with the same color scaling. 
 Comparing Figures \ref{fig:hrcorr}a, \ref{fig:hrcorr}b, and \ref{fig:hrcorr}c, it is clear that the
  correction procedure has recovered the true scatter appropriate to the A35 magnitudes, 
  and this scatter is now larger than that of the modern, and more precise, magnitudes. 
 Additionally, the subgiant sequence, shown in open boxes, is no longer detached from the 
  main sequence or giant branch, 
   instead a continuum of magnitudes is produced between these two phases of stellar evolution. 
 In this modified version of the A35 diagram, 
  the true transitionary nature of the subgiant sequences is revealed.
 Had it been realized at the time, it is possible that Gamow's conjecture 
  might have been voiced more vociferously in scholarly circles, and
  potentially the ultimate understanding of stellar evolution would have been hastened.

\section{Summary}\label{sec:c2sum}

   1. The discovery of subgiants culminated with the summary
paper by the Mount Wilson spectroscopists in 1935 that gave the
absolute magnitudes of 4179 stars determined empirically from line ratios in
their spectra that are sensitive to surface gravity.

   2. The 90 Mount Wilson subgiant candidates formed a distinct sequence
in the HR diagram between giant and dwarf stars and that was not
attached either to the main sequence or to the giant sequence (Figure \ref{fig:hr}).

   3. The problem posed at the 1957 Vatican Conference that high
weight trigonometric parallaxes for a few field subgiants put
them fainter than the subgiants in M67, believed then to be the
oldest stars in the Galaxy (along with the  globular clusters), was
solved with the later discovery of the older Galactic clusters of
NGC \,188 and NGC\,6791.

   4. A test is made of the reality of the $\sim$90 stars that the Mount Wilson spectroscopists
identified as part of the 
subgiant sequence via a star-by-star comparison of the Mount
Wilson and the \hip absolute magnitudes.  A majority of the 90 Mount Wilson
candidates that we have identified as the Adams et al. ``subgiants"
are clearly in the domain of subgiants when viewed in the modern \hip 
color-magnitude diagram.

5. A strong correlation is found between the A35 spectroscopically
derived absolute magnitudes for their subgiants and the corresponding absolute magnitudes
for these stars
derived using \hip data.  This statement holds even more broadly across
all luminosity classes in the A35 database.  
We find for the subgiants that
 $\langle M_{V, {\rm Mount Wilson}} - M_{V, {\rm \hip}}\rangle = -0.12 \pm 0.02$ 
  and with a dispersion of $1.02$ magnitudes, when restricted to stars having \hip $\sigma_{\pi}/\pi <  0.05$. 

   6. Tests for an observational selection
 incompleteness bias (e.g., the Malmquist bias) in the \citet{Stromberg30} calibration of the
 Mount Wilson subgiant sequence indicates that this bias did not artificially create
 the subgiant sequence in the A35 database as had been posited in the literature. 

  7. The reason for the narrow, disconnected A35 subgiant sequence
  is instead explained by 
  a subtle bias in the A35 calibration curves that,
  while well known, could not be 
  corrected for in the case of evolved stars due to
  the limited number of these stars in pre-\hip trigonometric parallax samples.
  However, even after use of the \hip parallaxes to correct for this historically
  well-discussed calibration bias
  we find that the A35 luminosity classes are robust and that 
  the detection --- and therefore discovery --- of subgiants by the Mount
  Wilson spectroscopists is upheld.
  
\section{A Note on the Preparation of this Manuscript}
This work began as a collaboration between the authors in 2005 and 
 progressed primarily by exchange of letters between AS and SRM until the death of AS in 2010.  
AS conceived the original concept and framework of the paper, and authored the original drafts of the manuscript.  
The historical insights described herein would not have been possible without AS's encyclopedic knowledge of,
 and indeed direct involvement in, many of the events described.  
Quantitative analyses and later stages in the crafting of the manuscript were carried on by RLB and SRM
 (in large part well after the passing of AS), 
 and were only possible after intense study of pertinent writings of the numerous historical characters 
 that played a role in the story described here. 
Special consideration was given to the handwritten letters, annotated drafts of the manuscript, 
 and relevant published papers by AS
 himself so that we might attempt the closest fidelity of the final work to his original intent.  
Much as described by Sandage after the death of Edwin Hubble 
 (albeit with far less significance of the product offered in the present paper!), 
 RLB and SRM found themselves 
 ``lost in the shadow of a giant" in attempting to bring this work to the conclusion originally 
 envisioned by the lead author.  
 Much of his original outline and text has been preserved in this edition of the manuscript, 
 but unfortunately we had to resort to our own less elegant prose in most of the later sections.
AS is rightfully maintained as the primary author in that he was the originator and true leader of the project, 
 but most especially to recognize his enthusiasm to rectify the historical judgment of the A35 work 
 and to uphold the reputation of the Mount Wilson spectroscopists.  
This enthusiasm remained vibrant even as AS's health began to fail in 2008, 
 and steadfast into his final months during which his production efficiency still well-outpaced 
 that of his coauthors (who, at the time, were well behind in completing their contributions).  
An edited draft of the manuscript, with his sections already long completed, 
  and a handwritten note signing off from the project due to declining health (dated August 7, 2010) 
  were sent directly by AS via postal delivery and received by SRM two days after ``Uncle Allan's" passing.

\section{Acknowledgements}
We wish to recognize helpful conversations with Richard Patterson and John Grula, 
 as well as highly insightful comments from the anonymous referee that led us 
  down a path to greatly improve this manuscript.
RLB acknowledges assistance with OCR software from the Alderman Library Scholars Lab at the University of Virginia.
Additionally, we acknowledge the encouragement from the family, friends
 and colleagues of AS, who also wished to see this work completed, including Mary Sandage, wife
 of the late Allan Sandage. 
RLB acknowledges support from the Jefferson Scholars Foundation as the C. Mark Pirrung Family Graduate Fellow in Astronomy
 and support from the Office of the Vice President of Research at the University of Virginia.
RLB and SRM acknowledge support from NSF grants AST-0407207, AST-0807945 and AST-1109718.

\appendix 
\section{APPENDIX A:~Estimation of Lutz-Kelker Corrections}\label{sec:app1}

  As briefly described in the main text, 
   the Lutz-Kelker Effect (LKE, hereafter) is a bias that occurs due to the finite magnitude 
   and sensitivity limits of trigonometric surveys. 
  The scattering of stars in and out of the sample due to observational errors is asymmetric,
   more stars are scattered into the survey limits due to the larger volume element just outside of the survey
   than are scattered out from the volume element just inside.
  In this appendix, we supplement the brief discussion of the LKE in the main text by providing 
   additional background on the LKE and a rigorous estimation of the appropriate offsets for the A35 catalog. 
  As stated in the main text from a qualitative assessment, we find the LKE to be a very small effect for the A35 sample,
   especially when appropriately restricted by $\sigma_{\pi}/\pi$.

 \subsection{Background}	
 
  \citet{Lutz73} provided the first mathematical derivation of the Lutz-Kelker Effect (LKE), which 
   had been alluded to in earlier work, 
   in particular by \citet[][but see Sandage \& Saha (2002) for a detailed history of the bias]{Trumpler53}.
  In this landmark paper, Lutz and Kelker found that the LKE was present for parallaxes independent 
   of apparent magnitude and that it was best characterized 
   by the dimensionless parameter, $\sigma_{\pi}/\pi$.
  The authors estimated the magnitude of the LKE as a function of $\sigma(\pi)/\pi$
   for the case of a uniform stellar distribution ($P(\pi) \sim \pi^{4}$) for observations 
   of stars with known intrinsic luminosity. 
 \citet{Hansons79} derived a generalized form of the Lutz \& Kelker derivation to other stellar 
   distributions ($P(\pi) \sim \pi^{n}$), but still did not account for targeting bias
   in the trigonometric survey.
 \citet{Lutz79} expanded the results of \citet{Lutz73} further and found two, as yet, 
   unexplored complications for generalizing the LKE to other datasets.
  First, the intrinsic luminosity function of the sample 
   must be known and {\it explicitly} included 
   in the derivation of the Lutz-Kelker correction.
  Second, the behavior of the LKE with $\sigma_{\pi}/\pi$ is more complex
   when an apparent magnitude ($m_{app}$) limit is imposed on the observational sample. 
  Lutz (1979) highlights the most insidious aspect of the LKE; 
   namely, that it has no general form.
  Instead, the appropriate Lutz-Kelker corrections must be derived 
    independently for the specifications of a sample or even a subsample within
    a larger sample already characterized for LKE.
  We note that there is an on-going debate in the literature as to both the nature and applicability
   of this bias, 
   in particular \citet{Smith03} and \citet{Francis14} present arguments for reconsidering
   the most common uses of the LKE.

\begin{figure}
 \includegraphics[width=\textwidth]{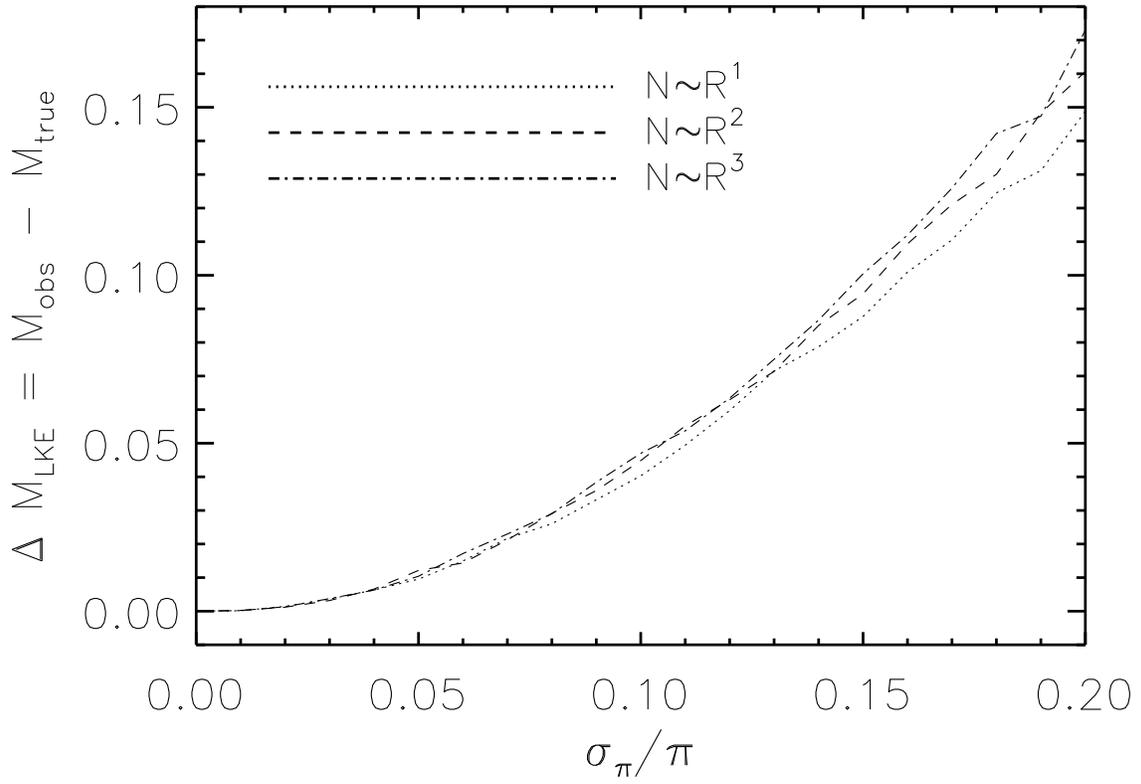}
\caption{\label{fig:lke} The Lutz-Kelker mean magnitude offset as a function of $\sigma_{\pi}/\sigma$. 
   These values were derived by adapting the methodology of S02 to properties of the A35 sample.
   The offset is given for three assumed stellar distributions, $N \sim R^{1}, R^{2}, R^{3}$ in 
    dotted, dashed and dot-dashed lines respectively. 
   Over all the corrections for each stellar distribution track each other well in the A35 volume. 
  }
\end{figure}

\subsection{Estimation of LKE for a General Sample}

  \citet[][S02, hereafter]{Sandage02} formalized the aforementioned complications into an easily adapted set of Monte-Carlo simulations 
   that permits the calculation of Lutz-Kelker corrections for any trigonometric sample. 
  This empirical approach is robust to concerns over the origin or functional form of the bias
   under debate in the literature.
  This work vividly illustrated the interplay of an $m_{app}$ limit 
   and the underlying stellar distribution in determining both the magnitude {\it and sign} of the Lutz-Kelker corrections.
  More specifically, their simulations revealed that the sign of the LKE can change 
   across $\sigma_{\pi}/\pi$ bins in a sample.
  In total, the complications explored by S02 emphasized the complexity 
   of the bias and how its behavior is intrinsically unintuitive. 
  To assess the role of the LKE on the \hip absolute magnitudes 
   matched to the A35 sample, we adapt the methodology of S02 to
   estimate the appropriate LKE corrections for the A35. 

  S02 was specifically focused on understanding the effect of the LKE on RR Lyrae
   stars, a class of stars typified by a narrow range of intrinsic luminosities.  
  Our goal of understanding the effect of the LKE on the \hip absolute magnitudes for the
   full A35 sample is more complicated due to the large range of intrinsic luminosities 
   in the A35 sample. 
  For the purposes of our calculation, we take the A35 absolute magnitudes to be the ``true'' 
   intrinsic magnitude of the star.
  Then, we generate simulated datasets that we use to explore the LKE for A35.
  
 To generate our simulated datasets, 
   we populate a spherical volume of radius, $R = 400$ pc, assuming a model stellar distribution.
 This radius includes the majority of the A35 sample as was demonstrated in Figure \ref{fig:spaenhauer}. 
 Stars are drawn from the A35 uniformly and, for full exploration of the bias, 
  we repeatedly select stars until we have $N = 100,000$ stars.
 Each star is randomly assigned a distance within our volume drawn from the model distributions.
 This process is completed independently for each of the three model stellar distributions 
  of S02, more specifically $N \sim R^{n}$ for $n = 1, 2, 3$. 
 After placing the star in our volume, we compute apparent magnitude 
  and the true parallax for each star, given its simulated distance. 
 The true parallax is then given a random error drawn from a Gaussian distribution with 
  width, $\sigma_{\pi}(m_{app})$.
 In a true parallax sample, this error will depend on the apparent magnitude of the star ($m_{app}$).  
 To simulate the errors from the \hip sample, we fit a linear relationship to the 
  rms errors in the A35 sample, resulting in:
    \begin{equation}
      \label{eq:hiperr}
      \sigma_{\pi}(m_{app}) = 0.07*m_{app} + 0.431. 
    \end{equation} 
 This final parallax, the true parallax with an added error based on the apparent magnitude, 
  is our simulated observed parallax for each star in our simulated sample.

 Having calculated an observed parallax, we can compute the implied absolute magnitude for 
  our simulated sample and compare it to the true magnitude used at the onset of the simulation.
 We calculate $\Delta M_{LKE} = M_{obs} - M_{true}$ for each star and compute the average value for the 
  full sample as a function of $\sigma_{\pi}/\pi$.

 The resulting Lutz-Kelker corrections are shown in Figure \ref{fig:lke} as a function of 
  $\sigma_{\pi}/\pi$ for each of the stellar distributions, $N \sim R^{n}$ with $n = 0, 1, 2$.
 On the whole, these corrections are negligible for $\sigma_{\pi}/\pi < 0.05$. 
  and, in general, are of the order of the errors typical of the A35 apparent magnitudes.   
 For a more specific comparison, we give the corrections for the $\sigma_{\pi}/\pi$ bins 
  used for the paper in Table 2 for each of the stellar distributions. 
 Thus, after proper consideration of the LKE, we conclude that it does not
  strongly influence our comparison of the A35 identified subgiants

\section{APPENDIX B: HOW RANDOM ERRORS REDUCED SCATTER IN THE A35 MAGNITUDE CALIBRATIONS}

 In the main text, we presented a qualitative description of the bias introduced by
  the A35 spectroscopists by the due to choices in their calibration techniques. 
 The bias creates two abnormal characteristics in the A35 color magnitude diagram (Figure 6a) 
  as compared to the \hip version (Figure 6c); these are:
 (i) compressed scatter along each luminosity sequence and 
 (ii) enhanced separation between sequences in on the luminosity axis.
 We applied a correction 
  procedure to correct for the bias, following the example of 
 \citet{Russell38,Russell40}, and produce the A35 HR diagram with its true observational
  and intrinsic scatter. 
 In supplement to the qualitative discussion given in the main text, we present 
  a mathematical description of the bias in the A35 diagram due to the 
  ``impartial'' calibration technique. 
 This mathematical approach is based on the summary given by \citet{Blaauw63} 
  of the various discussions from the literature 
  \citep[e.g.,][]{Stromberg39, Russell38, Russell40}
  and is translated into the modern lexicon for the benefit of the reader. 

 In general, luminosity calibration occurs in two stages. 
 First, a luminosity sensitive observable is identified, for A35
  a set of spectral line intensities ($I$) correlated with stellar luminosity.
 Second, an empirical relationship is measured to map between the luminosity 
  sensitive parameter and the absolute magnitude ($M$).
 To understand the origin of the systematic errors, we must consider how
  the magnitudes were calibrated. 

\subsection{The Ideal Case}
 First, we start with the ideal situation in which our line intensity distribution is well sampled
  by our observational data, meaning that the mean value ($I_{o}$) and dispersion ($\sigma_{I_{o}}$)
  calculated from the observational data are representative of their true, physical values.
 We start with a distribution of line intensities, $D(I)$, akin to the demonstration 
  of Figure \ref{fig:calibmeth}b in the main text. 
 For simplicity, we assume that $D(I)$ is a Gaussian distribution, 
  though the distribution could be any function.
 The mean value of the intensities is,
 \begin{equation} \label{eq:idealmean}
   \langle D(I) \rangle = I_{o}
 \end{equation}
  with a dispersion of $\sigma_{I}$.
 We require that $D(I)$ is uniquely connected to a distribution of intrinsic magnitudes, $D(M)$. 

 Again, we assume that the desired magnitude (luminosity) distribution, $D(M)$, will also be Gaussian,  
  such that the mean is given $\langle D(M) \rangle = M_{o}$ with a dispersion of $\sigma_{M}$. 
 We further assume that the relation to connect $D(M)$ to $D(I)$ is unique, 
  and, in imitation of the A35 calibration technique, that this relation is linear of the form:  

 \begin{equation} \label{eq:calib}
  M = \alpha I + \beta .
 \end{equation}

 By propagation of errors through Equation \ref{eq:calib},
  errors in $I$ ($\delta I$) translate to errors in $M$ ($\delta M$) by 

 \begin{equation} \label{eq:err}
   \delta M = \alpha \delta I .
 \end{equation}

 Likewise, the dispersion in $M$ due to uncertainty in $I$ is estimated by 

  \begin{equation} \label{eq:idealmagsigma}
    \sigma_{M} = \alpha \sigma_{I},
  \end{equation}

  and the mean magnitude of the resulting magnitude distribution $D(M)$ is
   determined by  
  \begin{equation} \label{eq:idealmagmean}
    M_{o} = \alpha I_{o} + \beta . 
  \end{equation}

 We choose from our line intensity sample, $D(I)$, a subset with luminosity classifications
  from an independent technique, ideally absolute magnitudes determined from trigonometric parallaxes. 
 The line intensity measurements ($I$) for this subsample 
  will be observed to cluster narrowly around some central value, $I_{1} \pm \Delta I$,
  and, ideally, this central value is $I_{o}$ and the spread ($\Delta I$) is given by $\sigma_{I}$
  (e.g., $I_{1} = I_{o}$ and $\Delta I\approx\sigma_{I}$). 
 For these stars we fit for $\alpha$ and $\beta$ as described in Equation \ref{eq:idealmean}, 
  and derive our set of absolute magnitudes ($D(M)$) with which we proceed to 
  perform our science experiment. 
 The mean values and dispersions of our absolute magnitude sample are computed
  directly from our line intensities following the relations given in 
  Equations \ref{eq:err}, \ref{eq:idealmagsigma}, and \ref{eq:idealmagmean}, respectively.

\begin{figure*}
 \includegraphics[width=\textwidth]{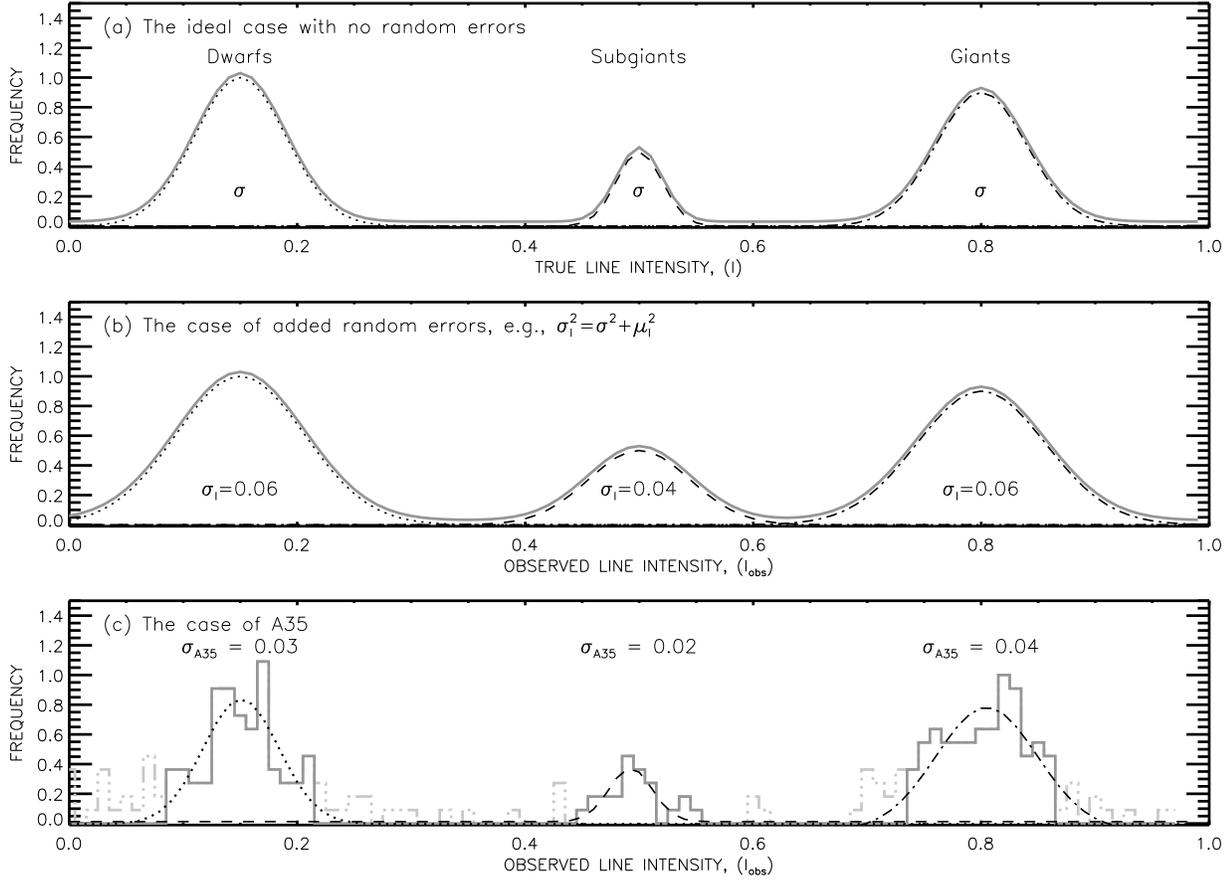}
 \caption{ \label{fig:distribution} 
  Schematic demonstration of the line intensity calibration scenarios described in the text.
  Each panel presents both a ``total'' distribution in thick grey, and 
   the distribution for each of the three component luminosity classes in A35: dwarfs (dotted),
   subgiants (dashed), and giants (dot-dashed). 
  In panel a, an ``ideal'' case is displayed with large samples of each class
   and no observational errors. 
  Here each distribution is easily identified and distinguished for calibration.
  In panel b, a random error, $\mu$, broadens each distribution, but the
   three distributions are still easily discriminated and their mean value is preserved.
  In panel c, we simulate the actual line intensity data of A35 by discretely
   sampling each of the luminosity class distributions (proportional to their
   representation in A35), and adding noise from a uniform distribution.
  In panel c, the light grey dot-dash distribution is the original sample, and the thick
   grey is the resulting sample after rejecting outliers and isolating the peaks.
  As in panel b, the mean value of each peak is preserved, but the resulting
   dispersion is suppressed from the outlier rejection.
  Though, we note that if all points (not just those passing the outlier rejection)
   are used in to fit each of the three distributions,
   then the true intrinsic dispersion is recovered.
  }
\end{figure*}

 This calibration case is deemed ``impartial,'' because our line intensity data were grouped into 
  calibration subsamples using a system ``impartial'' to the particulars of our observational
  systematics and uncertainties embedded within our line intensity distribution.
 Figure \ref{fig:distribution}a is a visualization of this ``ideal'' scenario.
 A priori, we divide our distribution into luminosity classes (dwarf, subgiant, and giant)
  from the overall distribution and these classes,
  determined by an independent means, can be used to convert line intensities into
  absolute magnitudes for our full sample.

\subsection{The Case of A35}
 In reality, however, our measurements are always imperfect due to 
  (i) observational uncertainties and (ii) ``cosmic uncertainties,'' (i.e.~intrinsic
  star to star variation) both of which behave as sources of ``random'' uncertainty in our dataset. 
 For sufficiently large sample sizes, these uncertainties can be minimized
  or, at least well characterized by the observational data itself
 (albeit they may not be readily distinguishable).
 {\it If} the the random uncertainties are well understood and appropriately characterized
  using independent datasets, 
  then the case of calibration reduces to that of the ``ideal''. 
  
 In practice, these two independent types of error -- observational (random) and cosmic --- are indistinguishable without 
  the acquisition additional data,
  and for our purpose here, we group them as a single, random error, $\mu_{I}$. 
 Now, we consider the effect of these, unrealized, random errors on the ideal 
  case presented previously. 
 This treatment is akin to that followed by the Mount Wilson spectroscopists for 
  the absolute magnitudes presented in A35, 
  and will demonstrate how incomplete consideration of random errors in 
  the observational data, here $D(I)$, can produce reduced 
  systematic errors in the resulting absolute magnitude distribution, $D(M)$. 

 For clarity, we redefine the line intensity distribution described previously, $D(I)$, to reflect
  observational error, $D(I_{obs})$.
 We assume it is Gaussian that can be described by $\langle D(I_{obs}) \rangle = I_{o,obs}$ with 
  scatter $\sigma_{obs}$, analogous to that case previously discussed. 
 The mean value is not affected by random errors, so
 
\begin{equation} \label{eq:a35mean}
  \langle D(I) \rangle = \langle D(I_{obs}) \rangle = I_{o},
 \end{equation}

 as was given in Equation \ref{eq:idealmean} for the ideal case. 
 The scatter in our line intensity distribution ($D(I)$), however, increases such that the uncertainties add in quadrature, i.e.,

 \begin{equation} \label{eq:totalerror}
  \sigma_{I,obs}^{2} = \sigma_{I}^{2} + \mu_{I}^{2}, 
 \end{equation}

 where $\sigma_{I}$ the scatter for the ideal case and $\mu_{I}$ is that from the observational
 uncertainty.
 This discussion is visualized in Figure \ref{fig:distribution}b, in which the
  ideal distributions of Figure \ref{fig:distribution}a are inflated with an additional random 
  uncertainty, $\mu$.

 As before, we proceed to determine a relationship between $D(I)$ and $D(M)$ for physically motivated
  subsets of $D(I)$, the spectral and (ultimately) luminosity classes of the target stars.
 This is accomplished by selecting appropriate stars within a narrow range 
  of some value $I_{1,obs} \pm \Delta I/2$ for each of our luminosity classes.
 Instead of grouping from an independent source as was done in the ideal case (Figure \ref{fig:distribution}a), we instead 
  plot a marginal distribution of all of our line intensities 
  and identify peaks in the resulting distribution.
 This distribution is represented in Figure \ref{fig:distribution}c by the dot-dash grey histogram, and
  we are able to identify the appropriate peaks, and their mean values, from this this distribution.
 We proceed to reject individual data points that seem distant from our mean values to isolate each distribution. 
 The result of this process is shown in the grey thick line of Figure \ref{fig:distribution}c. 
  By following this method, we make an assumption (consciously or unconsciously) 
   that the observed values do about some singular mean magnitude and that any scatter about that
   mean is completely driven by our observational errors. 
  In turn, we suppress intrinsic variation within the class, or at the very least, assume it is smaller
   than our observational uncertainty.
  Using independent classification metrics would have alleviated this concern. 
 
 Thus, we have now restricted our observational data to a narrow range around a mean
  and these values no longer properly sample either the observed scatter (Figure \ref{fig:distribution}b) 
  {\it or} the true scatter (Figure \ref{fig:distribution}a). 
 Stars with large deviations from the mean intensities are removed under the assumption that they are not related to the
  true distribution of $D(I)$  
  (i.e., due to some observational error and not due intrinsic variation within the class),
  and, as demonstrated in Figure \ref{fig:distribution}c, the resulting distributions are less broad. 
 In this instance, our intuitive procedure for propagating error, $\delta M = \alpha \delta I$,
  is no longer valid within these highly constrained sub-samples \citet{Trumpler53}
  and produces a constriction on the range of permissible values of $I$.

 The true values $I$ do not scatter symmetrically around the observed value $I_{1,obs}$. 
 Following \citet[][ \S 1.51]{Trumpler53}, for the case of ``accidental'' errors over a small range 
  in a univariate distribution, the true mean value, $\langle I \rangle$, is described by:\footnote{The 
   equation A8 comes from \S 1.51 {\it Correction of a univariate distribution for observational errors} 
   for the case of {\it Average errors for small intervals of the measured values} of \citet{Trumpler53}.}

 \begin{equation} \label{eq:expansion1}
  \langle I \rangle = I_{1,obs} + \frac{\mu_{I}^{2}}{D(I_{1,obs})} \frac{\delta D_{obs}}{\delta I_{obs}}(I_{1,obs}) + O(\mu_{I}^{2})
 \end{equation}
 Where $O(\mu_{I}^{2})$ represent terms of higher order in $\mu_{I}^{2}$ that are sufficiently small to ignore
  for our purposes.

 Since $D(I_{obs})$ is a Gaussian distribution, it and its derivative can be substituted into Equation A8,
  such that, with rearrangement, we obtain\footnote{This relation deviates 
  slightly from the precursor to Equation 7 in \citet{Blaauw63} due to what we suspect to be a typographical error 
  in that text.}:

 \begin{equation} \label{eq:expansion2}
   \langle I \rangle = I_{1,obs} - (I_{1,obs} - I_{o}) \times \frac{\mu_{I}^{2}}{\sigma_{I,obs}^{2}}.
 \end {equation}

 With additional rearrangement, we obtain a form that better elucidates the systematic error between 
  the true and the observed mean values: 

 \begin{equation} \label{eq:result} 
   \frac{\sigma_{I}^{2} + \mu_{I}^{2}}{\sigma_{I}^{2}} = \frac{(I_{1,obs}-I_{o})}{(\langle I \rangle - I_{o})}
 \end{equation}

 First, we note that the previous equation is the ratio of the relations, $(\langle I \rangle - I_{o})$ 
  and $(I_{1,obs}-I_{o})$ are, effectively, divisions on the true and observed
  intensity axis scales about the mean, respectively. 
 Second, the ratio of the true and observed axes is independent of the actual observed values ($I_{obs}$).
 Third, Equation \ref{eq:result} is always less than unity, as the observational uncertainty is always larger
  than zero ($\mu_{I} > 0$). 
 
 Now, we again fit a relationship to calibrate our subset, deriving:
  $ \langle M \rangle = \alpha \langle I \rangle + \beta$. 
 Though similar in form, this is {\it not} equivalent to the relationship derived earlier
  for our idealized case, $M_{o} = \alpha I_{o} + \beta$.
 Here, when we derive our calibration curve, we compare $\langle M \rangle$ against the 
  measured value, $\langle I_{obs} \rangle$, and not its true value, $I_{o}$.
 Effectively, as was demonstrated for $D(I_{obs})$ in Equation \ref{eq:result}, we measure our relation on an axis
  that is compressed by improper treatment of the observational uncertainty, 
  and this compression is propagated into the resulting distribution of absolute magnitudes $D(M_{obs})$.
 As demonstrated by Figure 9 in the main text, the resulting calibration curves are ``torqued''
  resulting in a narrower output distribution than is reasonable for the input observational data.

 Our derived values ($D(M_{obs})$) are also compressed compared to the true distribution $D(M)$. 
 There are two factors to consider in estimating the degree of compression in $D(M_{obs})$.
 First, we know that the observed dispersion is greater than the true dispersion 
  by a factor of $\sigma_{I,obs}/\sigma_{I}$, e.g. going from Figure \ref{fig:distribution}a and Figure \ref{fig:distribution}b.
 Second, the effective range of intensities is compressed by our outlier rejection
  by $\sigma^{2}_{I}/(\sigma^{2}_{I} + \mu^{2}_{I})$, e.g., the transition from 
  Figure \ref{fig:distribution}b and Figure \ref{fig:distribution}c.
 Combining the two factors, we arrive at the ratio of the true dispersion ($\sigma$)
  to the final dispersion of absolute magnitudes ($\sigma_{A35}$):

  \begin{equation} \label{eq:multiply}
   \frac{\sigma_{A35}}{\sigma} = \frac{\sigma_{I,obs}}{\sigma_{I}} \times \frac{\sigma^{2}_{I}}{(\sigma^{2}_{I} + \mu^{2}_{I})} 
   \end{equation}

 Substituting for $\sigma_{I,obs}$ from Equation \ref{eq:totalerror}, we obtain a final expression relating 
  the A35 dispersion to the true dispersion and random errors: 

  \begin{equation} \label{eq:scatter}
    \sigma_{A35} = \frac{\alpha \sigma}{\sqrt{1 + \mu_{I}^{2}/\sigma_{I}^2}}
  \end{equation}
 
 The denominator, $\sqrt{1 + \mu_{I}^{2}/\sigma_{I}^2}$, is always greater than unity, 
  and $\sigma_{A35}$ will, therefore, be smaller than the true dispersion ($\sigma$) 
  of the luminosity class. 
 Perhaps against our intuition, the resulting random errors in $D(M)$ are not increased by an amount 
  proportional to the uncertainty in the line intensities, 
  but are instead compressed by this observational error.
 Effectively, by removing outliers without insight from a means independent of our input data,
  we compress the effective output range of our absolute magnitude calibration.
 The compression in the output range explains both of the morphological concerns regarding
  the the shape and separation of the luminosity classes in the A35 CMD (Figure 6a).

\bibliographystyle{apj}
\bibliography{ms}

\begin{deluxetable}{l c | c c | l l | c c c | c c }
\tabletypesize{\tiny} \tablecaption{Data For the 90 Mount Wilson Subgiants That From Their Spectra Defined the Class
in 1935} \tablewidth{0pt}
\tablehead{ \colhead{Star Name} & \colhead{Star Name} &
 \colhead{\hip} & \colhead{\hip} & \colhead{Sp.Type} &
 \colhead{Sp.Type } & \colhead{$\pi$ (mas)} &
 \colhead{$\pi$ (mas)} &  \colhead{$\sigma(\pi) / \pi$} & \colhead{$M_V(\pi)$} & \colhead{$M_V(\pi)$}\\
 \colhead{Mt. W.} & \colhead{Hip.} & \colhead{$V$} &
 \colhead{$B-V$} & \colhead{Mt. W.} & \colhead{Hip.} & \colhead{Mt. W.} & \colhead{Hip.} & \colhead{Hip.} & \colhead{Mt. W.} & \colhead{
Hip.}}
\startdata
HD    28  & HIP   443  & 4.61 & 1.03 & K0 & K1    III    & 20 & 25.38 & 0.04 & 2.1 & 1.63  \\
HD  1037  & HIP  1176  & 6.63 & 1.03 & G8 & G8    III/IV & 13 &  7.76 & 0.10 & 2.4 & 1.08  \\
HD  2589  & HIP  2242  & 6.18 & 0.88 & G9 & K0    IV     & 17 & 25.37 & 0.02 & 2.5 & 3.20  \\
HD  3546  & HIP  3031  & 4.34 & 0.87 & G3 & G5    III    & 42 & 19.34 & 0.04 & 2.6 & 0.77  \\
HD  4398  & HIP  3607  & 5.49 & 0.98 & G6 & G8/K0 III    & 23 &  9.78 & 0.07 & 2.4 & 0.44  \\
HD  5268  & HIP  4257  & 6.15 & 0.91 & G3 & G5    IV     & 17 &  8.60 & 0.09 & 2.5 & 0.82  \\
HD  5286  & HIP  4288  & 5.46 & 1.01 & K1 & K1    IV     & 17 & 25.69 & 0.05 & 2.3 & 2.51  \\
HD  5395  & HIP  4422  & 4.62 & 0.96 & G4 & G8    III/IV & 36 & 15.84 & 0.04 & 2.6 & 0.62  \\ 
HD  6473  & HIP  5412  & 6.24 & 0.92 & G6 & K0           & 17 &  6.97 & 0.08 & 2.5 & 0.46  \\
HD  8512  & HIP  6537  & 3.60 & 1.07 & K0 & KO    III    & 46 & 28.48 & 0.03 & 2.1 & 0.87  \\
HD 10486  & HIP  8044  & 6.33 & 1.02 & K2 & K2    IV     & 13 & 18.04 & 0.04 & 2.0 & 2.61  \\
HD 16042  & HIP 12053  & 8.24 & 1.07 & G4 & K0    V      &  6 &  4.02 & 0.28 & 2.4 & 1.26  \\
HD 20618  & HIP 15514  & 5.91 & 0.86 & G5 & G8    IV     & 21 & 15.88 & 0.07 & 2.5 & 1.91  \\
HD 21467  & HIP 16181  & 6.03 & 0.95 & G6 & KO    IV     & 21 & 14.44 & 0.06 & 2.7 & 1.83  \\
HD 29613  & HIP 21685  & 5.46 & 1.05 & K1 & K0    III    & 21 & 16.42 & 0.04 & 2.2 & 1.54  \\ 
HD 34538  & HIP 24679  & 5.48 & 0.93 & G9 & G8    IV     & 21 & 20.69 & 0.04 & 2.3 & 2.06  \\
HD 34642  & HIP 24659  & 4.81 & 0.99 & K0 & K0/K1 III/IV & 30 & 29.63 & 0.02 & 2.3 & 2.17  \\
HD 37160  & HIP 26466  & 4.09 & 0.95 & G6 & G8    III/IV & 42 & 28.10 & 0.03 & 2.5 & 1.33  \\
HD 37601  & HIP 26941  & 6.05 & 0.95 & G9 & K0    III    & 17 & 16.26 & 0.04 & 2.3 & 2.11  \\
HD 37981  & HIP 26930  & 6.72 & 1.10 & K1 & K1    IV     & 10 &  8.98 & 0.11 & 2.0 & 1.49  \\
HD 39169  & HIP 27600  & 7.85 & 1.06 & K0 & G5           &  7 &  5.52 & 0.43 & 2.1 & 1.56  \\ 
HD 39364  & HIP 27654  & 3.76 & 0.98 & G7 & G8    III/IV & 79 & 29.05 & 0.02 & 3.4 & 1.08  \\
HD 40959  &\nodata     &\nodata&\nodata& G5 &\nodata     &  5 &\nodata&\nodata& 2.4 &\nodata \\ 
HD 45410  & HIP 31039  & 5.86 & 0.93 & G8 & K0    IV     & 20 & 17.56 & 0.04 & 2.5 & 2.08  \\
HD 46480  & HIP 31676  & 5.94 & 0.90 & G7 & G8    IV/V   & 24 & 18.82 & 0.04 & 2.9 & 2.31  \\
HD 55280  & HIP 35146  & 5.20 & 1.08 & K2 & K2    III    & 22 & 16.88 & 0.05 & 2.0 & 1.34  \\
HD 71952  & HIP 41894  & 6.23 & 1.01 & K0 & K0    IV     & 14 & 16.65 & 0.04 & 2.2 & 2.34  \\
HD 73593  & HIP 42604  & 5.35 & 0.99 & G6 & G0    IV     & 26 & 18.11 & 0.04 & 2.6 & 1.64  \\
HD 75558  & HIP 43463  & 7.38 & 0.91 & G3 & G5           & 11 &  5.37 & 0.20 & 2.5 & 1.03  \\
HD 77818  & HIP 44766  & 7.62 & 1.00 & K0 & K1    IV     &  8 &  8.72 & 0.10 & 2.2 & 2.32  \\
HD 78249  & HIP 44990  & 7.07 & 0.98 & K2 & K1    IV     & 10 & 15.11 & 0.05 & 2.2 & 2.97  \\ 
HD 79452  & HIP 45412  & 5.98 & 0.84 & G3 & G6    III    & 18 &  7.17 & 0.13 & 2.3 & 0.26  \\
HD 84406  & HIP 48034  & 6.94 & 0.95 & K0 & G5           & 13 & 13.24 & 0.05 & 2.5 & 2.55  \\
HD 84453  & HIP 47973  & 6.81 & 0.95 & K0 & K0    IV     & 11 & 12.06 & 0.07 & 2.1 & 2.22  \\
HD 86359  & HIP 48881  & 7.45 & 0.92 & G7 & G5           &  9 & 10.69 & 0.10 & 2.3 & 2.59  \\
HD 90752  & HIP 51204  & 7.23 & 0.97 & G9 & K0           &  9 &  7.27 & 0.20 & 2.2 & 1.54  \\
HD 91011  & HIP 51451  & 6.98 & 1.03 & K0 & K0           & 10 &  6.76 & 0.15 & 2.2 & 1.13  \\
HD 92588  & HIP 52316  & 6.25 & 0.88 & K1 & K1    IV     & 17 & 29.08 & 0.03 & 2.5 & 3.57  \\
HD 93636  & HIP 52882  & 6.15 & 1.14 & K1 & K0           & 14 &  5.45 & 0.14 & 2.0 &-0.17  \\
HD 94178  & HIP 53179  & 7.52 & 0.91 & G7 & G5           &  9 &  9.41 & 0.11 & 2.4 & 2.39  \\
HD 94264  & HIP 53229  & 3.79 & 1.04 & K2 & K0    III/IV & 46 & 33.40 & 0.02 & 2.2 & 1.41  \\
HD 96074  & HIP 54275  & 7.65 & 0.93 & G8 & G5           &  9 &  6.06 & 0.13 & 2.4 & 1.56  \\
HD 96436  & HIP 54336  & 5.52 & 0.96 & G7 & G9    IIICN  & 24 & 16.05 & 0.06 & 2.6 & 1.55  \\
HD 97100  & \nodata    &\nodata&\nodata& G5 &\nodata     &  5 &\nodata&\nodata& 2.5 &\nodata \\
HD 102928 & HIP 57791  & 5.62 & 1.06 & K0 & K0    IV     & 17 & 12.59 & 0.08 & 2.2 & 1.12  \\
HD 105639 & HIP 59285  & 5.95 & 1.12 & K3 & K3    III    & 14 & 11.94 & 0.07 & 1.8 & 1.34  \\
HD 110646 & HIP 62103  & 5.91 & 0.85 & G4 & G8    IIIp   & 22 & 14.26 & 0.05 & 2.8 & 1.68  \\
HD 111028 & HIP 62325  & 5.65 & 0.99 & K1 & K1    III/IV & 20 & 22.36 & 0.04 & 2.4 & 2.40  \\
HD 113817 & HIP 63960  & 7.09 & 0.98 & G8 & K0    III    & 11 &  4.67 & 0.20 & 2.4 & 0.44  \\
HD 115202 & HIP 64725  & 5.21 & 1.01 & K1 & K1    III    & 23 & 25.68 & 0.03 & 2.1 & 2.26  \\
HD 116713 & HIP 65535  & 5.11 & 1.18 & K1 & Kp           & 19 & 15.73 & 0.05 & 1.6 & 1.09  \\
HD 123409 & HIP 68955  & 6.89 & 1.00 & G6 & K0           & 13 &  5.71 & 0.14 & 2.5 & 0.67  \\
HD 126400 & HIP 70538  & 6.48 & 0.94 & G7 & K0    III    & 14 & 12.99 & 0.08 & 2.4 & 2.05  \\
HD 127243 & HIP 70791  & 5.58 & 0.86 & G4 & G3    IV     & 26 & 10.59 & 0.06 & 2.7 & 0.70  \\
HD 138716 & HIP 76219  & 4.61 & 1.00 & K1 & K1    IV     & 29 & 34.54 & 0.02 & 2.1 & 2.30  \\
HD 140301 & HIP 77007  & 6.30 & 1.12 & K0 & K0    III    & 13 &  7.94 & 0.11 & 2.0 & 0.80  \\
HD 140687 & \nodata    &\nodata&\nodata& K1 &\nodata     &  9 &\nodata&\nodata& 2.3 &\nodata \\
HD 142091 & HIP 77655  & 4.79 & 1.00 & K1 & K0    III/IV & 26 & 32.13 & 0.02 & 1.9 & 2.32  \\
HD 143586 & \nodata    &\nodata&\nodata& G9 &\nodata     &  5 &\nodata&\nodata& 2.3 &\nodata \\
HD 150275 & HIP 80850  & 6.35 & 1.00 & K0 & K1    III    & 14 &  8.00 & 0.07 & 2.2 & 0.87  \\
HD 151216 & HIP 81977  & 9.17 & 1.03 & K1 & K2           &  4 &  3.20 & 0.36 & 2.2 & 1.70  \\
HD 152781 & HIP 82861  & 6.33 & 0.95 & K2 & K0/K1 III/IV & 14 & 24.72 & 0.03 & 2.3 & 3.30  \\
ADS 10394B & \nodata   &\nodata&\nodata& K0 &\nodata     &  5 &\nodata&\nodata& 1.8 &\nodata \\
HD 156461 & HIP 84647  & 7.23 & 1.02 & G3 & G8    III    & 11 &  5.86 & 0.20 & 2.5 & 1.07  \\
HD 159466 & HIP 85979  & 6.52 & 0.95 & G4 & G8    III    & 14 &  6.40 & 0.15 & 2.5 & 0.55  \\
HD 160042 & HIP 86352  & 6.19 & 0.83 & G7 & G6    III/IV & 14 &  9.66 & 0.09 & 2.4 & 1.11  \\
HD 162076 & HIP 87158  & 5.69 & 0.94 & G5 & G5    IV     & 20 & 13.04 & 0.05 & 2.3 & 1.27  \\
HD 165438 & HIP 88684  & 5.74 & 0.97 & K1 & K1    IV     & 17 & 28.61 & 0.03 & 2.1 & 3.02  \\
HD 167042 & HIP 89047  & 5.97 & 0.94 & K0 & K1    III    & 18 & 20.00 & 0.03 & 2.2 & 2.48  \\
HD 173399 & HIP 91782  & 8.99 & 0.42 & G2 & F2    V      & 12 &  7.39 & 0.89 & 2.5 & 3.33  \\
HD 173949 & HIP 91915  & 6.02 & 0.97 & G7 & G7    IV     & 16 &  8.95 & 0.06 & 2.3 & 0.78  \\
HD 181391 & HIP 95066  & 4.98 & 0.94 & K0 & G8    III/IV & 24 & 21.17 & 0.04 & 2.3 & 1.61  \\
HD 185351 & HIP 96459  & 5.17 & 0.93 & K0 & K0    III    & 27 & 24.64 & 0.02 & 2.4 & 2.13  \\
HD 194433 & HIP 100852 & 6.24 & 0.96 & K1 & K1    IV     & 17 & 25.21 & 0.05 & 2.5 & 3.25  \\
HD 196925 & HIP 101082 & 5.96 & 0.94 & G8 & K0    III+   & 18 & 15.88 & 0.03 & 2.4 & 1.96  \\
HD 198149 & HIP 102422 & 3.41 & 0.91 & G7 & K0    IV     & 63 & 69.73 & 0.01 & 2.6 & 2.63  \\
HD 198732 & HIP 103071 & 6.32 & 0.88 & G5 & K0    III    & 19 & 13.33 & 0.09 & 2.6 & 1.94  \\
HD 198896 & HIP 102990 & 7.91 & 0.62 & G7 & Am           &  9 &  0.33 & 6.55 & 3.3 &-4.50  \\
HD 199223 & HIP 103301 & 6.04 & 0.82 & G6 & G6    III/IV & 17 &  9.02 & 0.13 & 2.3 & 0.82  \\
HD 200004 & HIP 103728 & 6.55 & 0.85 & G3 & G6/G8 III    & 14 &  6.05 & 0.16 & 2.3 & 0.46  \\
HD 202403 & HIP 104844 & 7.08 & 0.80 & G5 & G5           &  8 &  5.21 & 0.23 & 2.3 & 0.66  \\
BD +15 466 &\nodata    &\nodata&\nodata& K3 &\nodata     &  4 &\nodata&\nodata& 2.1 &\nodata\\
HD 212943 & HIP 110680 & 4.78 & 1.04 & K0 & K0    III    & 30 & 20.39 & 0.04 & 2.3 & 1.33  \\
HD 216640 & HIP 113148 & 5.53 & 1.11 & K4 & K1    III    & 17 & 23.27 & 0.03 & 1.9 & 2.36  \\
HD 218527 & HIP 114273 & 5.42 & 0.91 & G4 & G8    IV     & 23 & 11.64 & 0.11 & 2.4 & 0.75  \\
HD 221148 & HIP 115953 & 6.26 & 1.12 & K3 & K3    IIIvar & 13 & 20.44 & 0.04 & 2.0 & 2.81  \\
HD 221639 & HIP 116251 & 7.20 & 0.92 & G9 & K1    V      & 10 & 14.58 & 0.05 & 2.4 & 3.02  \\
HD 222107 & HIP 116584 & 3.81 & 0.98 & G7 & G8    III/IV & 40 & 38.74 & 0.02 & 2.3 & 1.75  \\
HD 222404 & HIP 116727 & 3.21 & 1.03 & K1 & K1    IV     & 55 & 72.50 & 0.01 & 2.1 & 2.51  
\enddata
\end{deluxetable}

\begin{deluxetable}{c c c c}
\tablecaption{Lutz-Kelker Corrections for A35 Sample}
\label{table:lke}
\tablehead{\colhead{$\sigma_{\pi}/\pi$} & \colhead{$N\sim R^{1}$} & \colhead{$N\sim R^{2}$} & \colhead{$N\sim R^{3}$} \\
  \colhead{ } & \colhead{$\Delta M_{LKE}$} & \colhead{$\Delta M_{LKE}$} & \colhead{$\Delta M_{LKE}$} }
 \startdata
 0.01 & 0.000 & 0.000 & 0.000 \\
 0.05 & 0.095 & 0.012 & 0.010 \\
 0.10 & 0.040 & 0.045 & 0.047 \\
 0.15 & 0.088 & 0.094 & 0.100 \\
 0.20 & 0.149 & 0.161 & 0.173 \\
 0.25 & 0.205 & 0.222 & 0.241  
\enddata
\end{deluxetable}

\end{document}